%% file: author.tex
\documentclass[graybox]{svmult}

\usepackage{mathptmx}       
\usepackage{helvet}         
\usepackage{courier}        
\usepackage{type1cm}        
\usepackage{makeidx}         
\usepackage{graphicx}        
\usepackage{multicol}        
\usepackage[bottom]{footmisc}

\makeindex             

\newcommand{\supervisory}{SPECTR}
\newcommand{\multiPow}{MM-Pow}
\newcommand{\multiPerf}{MM-Perf}
\newcommand{\fs}{FS}

\usepackage{soul}

\usepackage{booktabs}   
\usepackage{multirow}
\usepackage[caption=false]{subfig}

\usepackage{tabularx}
\usepackage{wrapfig}
\usepackage{makecell}
\usepackage{amssymb}
\usepackage{algorithm, algorithmic}
\usepackage[numbers]{natbib}
\usepackage{tikz}
\usetikzlibrary{arrows.meta} 
\usetikzlibrary{positioning, calc, fit}
\usetikzlibrary{shapes,shapes.callouts}
\usetikzlibrary{decorations.markings}
\usetikzlibrary{automata}

\pgfdeclarelayer{bg1}    
\pgfdeclarelayer{bg2}    
\pgfdeclarelayer{vg1}    
\pgfsetlayers{bg2,bg1,main,vg1}  
\usepackage{pgfplots,pgfplotstable}
\pgfplotsset{compat=1.14}
\pgfplotsset{
	/pgfplots/ybar legend/.style={
		/pgfplots/legend image code/.code={%
			\draw[##1,/tikz/.cd,yshift=-0.25em]
			(0cm,0cm) rectangle (3pt,0.8em);},
	},
}

\usepackage{adjustbox}

\usepackage{color,colortbl}

\usepackage{stackengine}

\usepackage{siunitx}

\begin{document}

\title*{Intelligent Management of Mobile Systems through Computational Self-Awareness}
\author{Bryan Donyanavard, Amir M. Rahmani, Axel Jantsch, Onur Mutlu, and Nikil Dutt}
\authorrunning{Donyanavard, Rahmani, Jantsch, Mutlu, Dutt}
\institute{Bryan Donyanavard \at UC Irvine, \email{bdonyana@uci.edu}
\and Amir M. Rahmani \at UC Irvine \email{amirr1@uci.edu}
\and Axel Jantsch \at TU Vienna \email{axel.jantsch@tuwien.ac.at}
\and Onur Mutlu \at ETH Zurich \email{onur.mutlu@inf.ethz.ch}
\and Nikil Dutt \at UC Irvine \email{dutt@uci.edu}}
%
%
\maketitle

\abstract*{
Runtime resource management for many-core systems is increasingly complex. 
The complexity can be due to diverse workload characteristics with conflicting demands or limited shared resources such as memory bandwidth and power. 
Resource management strategies for many-core systems must distribute shared resource(s) appropriately across workloads, while coordinating the high-level system goals at runtime in a scalable and robust manner.\newline\indent
To address the complexity of dynamic resource management in many-core systems, state-of-the-art techniques that use heuristics have been proposed. 
These methods lack the formalism in providing robustness against unexpected runtime behavior. One of the common solutions for this problem is to deploy classical control approaches with bounds and formal guarantees. 
Traditional control theoretic methods lack the ability to adapt to (1) changing goals at runtime (i.e., \emph{self-adaptivity}), and (2) changing dynamics of the modeled system (i.e., \emph{self-optimization}).\newline\indent
In this chapter, we explore adaptive resource management techniques that provide self-optimization and self-adaptivity by employing principles of computational self-awareness, specifically \emph{reflection}. 
By supporting these self-awareness properties, the system will reason about the actions it takes by considering the significance of competing objectives, user requirements, and operating conditions while executing unpredictable workloads.
}

\abstract{
Runtime resource management for many-core systems is increasingly complex. 
The complexity can be due to diverse workload characteristics with conflicting demands, or limited shared resources such as memory bandwidth and power. 
Resource management strategies for many-core systems must distribute shared resource(s) appropriately across workloads, while coordinating the high-level system goals at runtime in a scalable and robust manner.\newline\indent
To address the complexity of dynamic resource management in many-core systems, state-of-the-art techniques that use heuristics have been proposed. 
These methods lack the formalism in providing robustness against unexpected runtime behavior. 
One of the common solutions for this problem is to deploy classical control approaches with bounds and formal guarantees. 
Traditional control theoretic methods lack the ability to adapt to (1) changing goals at runtime (i.e., \emph{self-adaptivity}), and (2) changing dynamics of the modeled system (i.e., \emph{self-optimization}).\newline\indent
In this chapter, we explore adaptive resource management techniques that provide self-optimization and self-adaptivity by employing principles of computational self-awareness, specifically \emph{reflection}. 
By supporting these self-awareness properties, the system can reason about the actions it takes by considering the significance of competing objectives, user requirements, and operating conditions while executing unpredictable workloads.
}

\section{Introduction}

Battery powered-devices are the most ubiquitous computers in the world. 
Users expect the devices to support high performance applications running on same device, sometimes at the same time.
The devices support a wide range of applications, from interactive maps and navigation, to web browsers and email clients.
In order to meet the performance demands of the complex workloads, increasingly powerful hardware platforms are being deployed in battery-powered devices.
These platforms include a number of configurable knobs that allow for a tradeoff between power and performance, e.g., dynamic voltage and frequency scaling (DVFS), core gating, idle cycle injection, etc.
These knobs can be set and modified at runtime based on workload demands and system constraints.
Heterogeneous manycore processors (HMPs) have extended this principle of dynamic power-performance tradeoffs by incorporating single-ISA, architecturally differentiated cores on a single processor, with each of the cores containing a number of independent tradeoff knobs.
All of these configurable knobs allow for a large range of potential tradeoffs.
However, with such a large number of possible configurations, HMPs require intelligent runtime management in order to achieve application goals for complex workloads while considering system constraints.
Additionally, the knobs may be interdependent, so the decisions must be coordinated.
In this chapter, we explore the use of computational self-awareness to address challenges of adaptive resource management in mobile multiprocessors.


\subsection{Computational Self-awareness}

Self-aware computing is a new paradigm that does not strictly introduce new research concepts, but unifies overlapping research efforts in disparate disciplines \cite{lewis2016self}.
The concept of self-awareness from psychology has inspired research in autonomous systems and neuroscience, and existing research in fields such as adaptive control theory support properties of self-awareness.
This chapter addresses key challenges for achieving computational self-awareness that can make the design, maintenance and operation of complex, heterogeneous systems adaptive, autonomous, and highly efficient.
Computational self-awareness is the ability of a computing system to recognize its own state, possible actions and the result of these actions on itself, its operational goals, and its environment, thereby empowering the system to become autonomous \cite{jantsch2017self}.
%
An infrastructure for system introspection and reflective behavior forms the foundation of self-aware systems. 

\subsubsection{Reflection}

\begin{figure}
    \centering
        \resizebox{\textwidth}{!}{\input{fig/feedback_controller_reflection.tex}}
    \caption{Feedback loop overview. The bottom part of the figure represents a simple observe-decide-act loop. The top part (in \textcolor{blue}{blue}) adds the reflection mechanism to this loop, enabling predictions for smart decision making.}
    \label{fig:FB-Controller}
\end{figure}
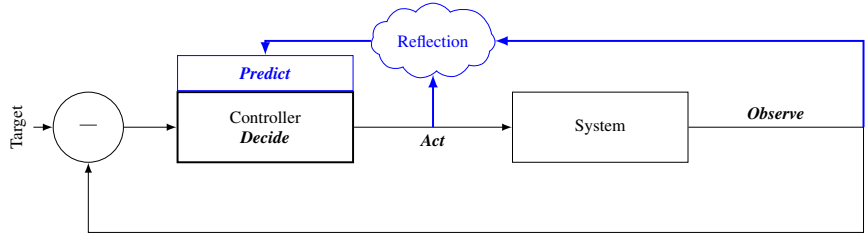

Reflection can be defined as \emph{the capability of a system to reason about itself and act upon this information} \cite{Smith1982}.
A reflective system can achieve this by maintaining a representation of itself (i.e., a self-model) within the underlying system, which is used for reasoning.
Reflection is a key property of self-awareness.
Reflection enables decisions to be made based on both \emph{past} observations, as well as predictions made from past observations.
Reflection and prediction involve two types of models: (1) a self-model of the subsystem(s) under control, and (2) models of other policies that may impact the decision-making process.
Predictions consider \emph{future} actions, or events that may occur before the next decision, enabling "what-if" exploration of alternatives. 
Such actions may be triggered by other resource managers running with a shorter period than the decision loop. 
The top half of Figure~\ref{fig:FB-Controller} (in \textcolor{blue}{blue}) shows prediction enabled through reflection that can be utilized in the decision making process of a feedback loop.
The main goal of the prediction model is to estimate system behavior based on potential actuation decisions.
This type of prediction is most often performed using linear regression-based models \cite{Muck2015,Pricopi2013,Annamalai2013,Singh2009} due to their 
simplicity, while others employ a binning-based approach in which metrics sensed at runtime are used to classify workloads into categories \cite{Liu2013,Donyanavard2016}.

\subsection{Closed-loop Resource Management in Mobile Systems}

Runtime resource management for many-core systems is increasingly challenging due to the complex interaction of:
i) integrating hundreds of (heterogeneous) cores and uncore components on a single chip, ii) limited amount of system resources (e.g., power, cores, interconnects), iii) diverse workload characteristics with conflicting constraints and demands, and iv) increasing pressure on shared system resources from data-intensive workloads.
As system size and capability scale, designers face a large space of configuration parameters controlled by actuation knobs, which in turn generate a very large number of cross-layer actuation combinations \cite{Zhang:2016:MPU:2872362.2872375}.
Making runtime decisions to configure knobs in order to achieve a simple goal (e.g., maximize performance) can be challenging.
That challenge is exacerbated when considering a goal that may change throughout runtime, and consist of conflicting objectives (e.g., maximize performance while minimizing power consumption).
Additionally, ubiquitous mobile devices are expected to be general-purpose, supporting any combination of applications (i.e., workloads) desired by users, often without any prior knowledge of the workload.

Designers face a large space of configuration parameters that often are controlled by a limited number of actuation knobs, which in turn generate a very large number of cross-layer actuation configurations. 
For instance, Zhang and Hoffman~\cite{Zhang:2016:MPU:2872362.2872375} show that for an 8-core Intel Xeon processor, combining only a handful of actuation knobs (such as clock frequency and Hyperthreading levels)  generates over 1000 different actuation configurations;
they use binary search to efficiently explore the configuration space for achieving a \emph{single} goal: cap the Thermal Design Power (TDP) while maximizing performance. 
Searching the configuration space is common practice in many similar single-goal, heuristic-based, runtime resource management approaches \cite{Raghavendra:2008:NPS,Choi:2006:LSP,Tembey:2010:CCR:2185870.2185902,Vega:2013:CUD,Cochran:2011:PCA:2155620.2155641}. 
While there is a large body of literature on ad-hoc resource management approaches for processors using heuristics and thresholds \cite{Deng:2012,Hwisung,Ebrahimi:2010,David:2011}, rules \cite{Isci:2006:AEM, Dhodapkar:2002}, solvers \cite{Petrica:2013,Hanumaiah:2014:SST}, and predictive models \cite{Bitirgen:2008:CMM,Dubach:2013:DMA,Dubach:2010:PMD:1934902.1934992,Donyanavard:2016},
there is a lack of formalism in providing guarantees for resource management of complex many-core systems.

Closed-loop systems have been used extensively to improve the state of a system by configuring knobs in order to achieve a goal. 
Closed-loop systems traditionally deploy an \emph{Observe}, \emph{Decide} and \emph{Act} (ODA) feedback loop (lower half (in black) of Figure~\ref{fig:FB-Controller}) to determine the system configuration.
In an ODA loop, the observed behavior of the system is compared to the target behavior, and the discrepancy is fed to the controller for decision making. 
The controller invokes actions based on the result of the \emph{Decide} stage. 

Resource management approaches in the literature can be classified into three main classes: 
(1) hueristic-based-approaches \cite{Petrica:2013,Hanumaiah:2014:SST,Mahajan:2016,Fu:2011,Teodorescu:2008,Sui:2016:PCA:2872362.2872402,7879873,Isci:2006:AEM, Dhodapkar:2002,Fan:2016:CSG:2954680.2872383,Bitirgen:2008:CMM,Dubach:2013:DMA,Dubach:2010:PMD:1934902.1934992,Deng:2012,Hwisung,Zhang:2016:MPU:2872362.2872375,Bartolini,Wang:2011,Wang:2009:TPC:1555754.1555794,Yan:2016,Lo:2015:PPT:2830772.2830776,Su:2014:prediction,Wang:2016:RTO:2954680.2872382,SomuMuthukaruppan:2014:PTB:2654822.2541974,Donyanavard:2016,Ebrahimi:2010,David:2011,Ebrahimi:2011,Das:2009,Das-HPCA,Chang:2017:URO:3107080.3084447,Subramanian:2015:ASM:2830772.2830803,Subramanian:2013},
(2) control-theory-based approaches \cite{5717893,Hoffmann:2011,MOC,MoC2,Mishra:2010:CCC,Wu:2005,Wu:2004:FOM:1037949.1024423,Ebrahimi:2009:CCM:1669112.1669154,Ma:2011,Muthukaruppan:2013:HPM:2463209.2488949,Kadjo:2015:CAE:2744769.2744773,Hoffmann:CoAdapt,Srikantaiah,Anil-ICCAD,Haghbayan-TC,MIMO-16}, and
(3) stochastic/machine-learning-based approaches \cite{7827638,Bitirgen:2008:CMM,Dubach:2010:PMD:1934902.1934992,Delimitrou:2014:QRQ:2541940.2541941,Ipek:2008:SMC:1381306.1382172}. 
Recent work has combined aspects of machine learning and feedback control \cite{Mishra2018}.
In addition, there have been efforts to enable coordinated management in computer systems in various ways~\cite{Bitirgen:2008:CMM,Raghavendra:2008:NPS,Choi:2006:LSP,IJES.2009,Juang:2005,Wu:2016,Tembey:2010:CCR:2185870.2185902,Vega:2013:CUD,Cochran:2011:PCA:2155620.2155641,Dubach:2010:PMD:1934902.1934992,Ebrahimi:2011,Ebrahimi:2010,David:2011,Das:2009,Stuecheli:2010,Lee:2010,Das:2010:AEP:1815961.1815976, Pothukuchi:2018}.
These works coordinate and control multiple goals and actuators in a non-conflicting manner by adding an ad-hoc component or hierarchy to a controller.

In this chapter, we demonstrate the effectiveness of computational self-awareness in adaptive resource management for mobile processors.
The self-aware resource managers discussed are implemented using classical and hierarchical control.

\section{Self-optimization}

Self-optimization is the ability of a system to adapt and act efficiently by itself in the face of \emph{internal stimuli}.
We consider internal stimuli as changes related to dynamics in the system's self-model, i.e., model inaccuracy.
Internal stimuli does not necessarily include workload itself, but if the self-model is application-dependent, workload changes may be the source of internal stimuli.
For example, if the system's self-model is application-dependent, and the executing application changes, a self-optimizing manager will have the ability to reason and act towards achieving the system goal(s) efficiently for the new application.
However, if the system's self-model is rigid and the system dynamics used to reason and act are oversimplified, model inaccuracies may lead to undesirable or inefficient decisions when the application changes.

\subsection{Background and Motivation}

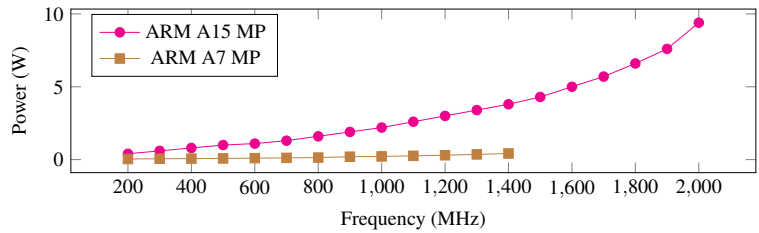
\begin{figure}[]
    \centering
        \input{fig/arm_big_little_cluster_f-p.tex}
    \caption{Cluster power vs. operating frequency (from \cite{Donyanavard2018b}).}
    \label{fig:motiv:f_p}
\end{figure}

Dynamic voltage/frequency scaling (DVFS) has been established as an effective technique to improve the power-efficiency of chip-multiprocessors (CMPs) \cite{herbert07}.
In this context, numerous closed-loop control-theoretic solutions for chip power management \cite{MOC,hoffmann2011dynamic,mishra2010coordinated,muthukaruppan2013hierarchical,ma2011scalable,wang2011adaptive} have been proposed. 
These solutions employ \emph{linear control} techniques to limit the power consumption by controlling the CMP operating frequency.
However, the relationship between operating frequency and power is often \emph{nonlinear}. 
Figure~\ref{fig:motiv:f_p} illustrates this by showing total power consumed by a 4-core ARM A15 cluster executing a CPU-intensive workload through its entire frequency range (200MHz--2GHz), along with the total power consumed by a 4-core ARM A7 cluster through its frequency range (200MHz--1400MHz).
While the A7 cluster frequency-power relationship is almost linear, the A15 cluster's larger frequency range (and more voltage levels) results in a nonlinear relationship.
Using a linear model to estimate the behavior of such a system leads to inaccuracies.
Inaccurate models result in inefficient controllers, which defeats the very purpose of using control theoretic techniques for power management.

Ideally, control-theoretic solutions should provide formal guarantees, be simple enough for runtime implementation, and handle nonlinear system behavior.
Static linear feedback controllers can provide robustness and stability guarantees with simple implementations,
while adaptive controllers modify the controller at runtime to adapt to the discrepancies between the expected and the actual system behavior. 
However, modifying the controller at runtime is a costly operation that also invalidates the formal guarantees provided at design time.

Instead, consider integrating multiple linear models within a single controller implementation in order to estimate nonlinear behavior of DVFS for CMPs.
This is a well-established and lightweight adaptive control theoretic technique called \emph{gain scheduling}.

\subsubsection{Classical Control}
\label{sec:background}

Discrete-time control techniques are the most appropriate to implement control of computer systems.
The proportional-integral-derivative (PID) controller is a simple and flexible classical feedback controller that computes control input $u(t)$ based on the error $e(t)$ between the measured output and reference output:    
\vspace{-1.0em}
\begin{equation}
	\vspace{-0.5em}
    u(k) = K_p e(k) + K_i \sum_0^k{e(k)\Delta t} + K_d\frac{\Delta	e(k)}{\Delta t}
\end{equation}
$K_p$, $K_i$, and $K_d$ are control parameters for the proportional, integral, and derivative gains respectively.

PI controllers\footnote{
Due to the significant stochastic component of computer systems, PI controllers are preferred over PID controllers \cite{Hellerstein:2004:FCC}.
} have been successfully used to manage DVFS of CMPs \cite{mishra2010coordinated,wu2004formal,muthukaruppan2013hierarchical,ma2011scalable,wang2011adaptive}.
Mishra et al.~\cite{mishra2010coordinated} propose the use of PID controllers for VF islands.
The authors model power consumption based on the assumption that the difference relationship between power consumption in successive intervals can be approximated linearly as a function of frequency, which only holds for limited range.
Similarly, Hoffman et al.~\cite{hoffmann2011dynamic} propose a feedback control technique for power management that includes DVFS, and their transfer function assumes a linear relationship between power and frequency.
However, Figure~\ref{fig:motiv:f_p} shows that $f \rightarrow P$ becomes nonlinear at higher frequencies.
Inaccuracies in linear estimation of nonlinear systems can negatively impact the steady-state error and transient response of the controller.
Take for example a system operating under a power budget, or experiencing a thermal emergency -- a DVFS controller designed from an inaccurate model could lead to wasted power or even unnecessary operation at an unsafe frequency.

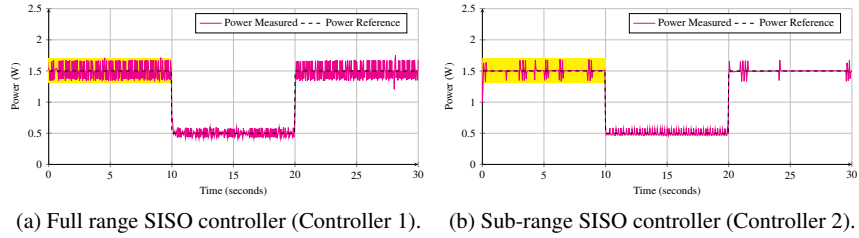
\begin{figure}[]
\centering
        \subfloat[Full range SISO controller (Controller 1).]{
            \resizebox{0.48\textwidth}{!}{\input{fig/gsc_motivation_time_full.tex}}
            \label{fig:motiv:full-siso-time}
        }%
        \subfloat[Sub-range SISO controller (Controller 2).]{
            \resizebox{0.48\textwidth}{!}{\input{fig/gsc_motivation_time_l2l3.tex}}
            \label{fig:motiv:L2L3-siso-time}
        }%
        \caption{Time plots of two DVFS controllers tracking a dynamic power reference (from \cite{Donyanavard2018b}).}
        \label{fig:motiv}
\end{figure}

Consider a DVFS controller for a 4-core CMP with a single frequency domain.
The first steps in designing a controller are defining the system and identifying the model. 
The power consumption of our CMP is not linear across the entire range of supported operating frequencies (200MHz--2GHz), which makes it challenging to model the entire range with a single linear estimation. 
However, we can divide the measured output (power) for the entire range of frequencies into multiple \textit{operating regions} that exhibit linear behavior.
In this example, we identify a model for two different systems: (1) the CMP's behavior through all operating frequencies; 
(2) the CMP's behavior through a sub-range of the operating frequencies.
This specific operating region spans the frequency sub-range of 200MHz--1200MHz.
Using these models, we can generate two different $f \rightarrow P$ Single-Input-Single-Output (SISO) PI controllers, and compare them using measured SASO analysis \cite{Hellerstein:2004:FCC}, focusing on \emph{Accuracy} and \emph{Settling time}. 
We refer to the full-range controller as Controller 1, and the sub-range controller as Controller 2.
Figure~\ref{fig:motiv} displays Controller 1 (Fig.~\ref{fig:motiv:full-siso-time}) and Controller 2's (Fig.~\ref{fig:motiv:L2L3-siso-time}) ability to track a dynamic power reference over time for our CMP.

\emph{Accuracy} is defined by the steady-state error between the measured output and reference input, e.g., the yellow highlighted region in Figure~\ref{fig:motiv:L2L3-siso-time} from 0-10 seconds. 
We calculate the steady-state error as the mean squared error (MSE) between the measured power and reference power.
Both controllers are able to track within 1\% of the target power.
However, the MSE of Controller 2 is 0.003, while that of Controller 1 is 0.013 -- an order of magnitude larger.
This byproduct of model inaccuracy translates into wasted power and undesirable operating frequency, as well as unnecessary changes in the frequency control input (i.e., increased control effort cost).

\emph{Settling time} is the time it takes to reach sufficiently close to the steady-state value after the reference values are specified, e.g., when the reference changes in Figure~\ref{fig:motiv:L2L3-siso-time} at 10 seconds.
The settling time of Controller 2 is $40ms$ on average, while Controller 1 is more than double on average at $100ms$.
Because our actuation periods are $50ms$, this means that our sub-range controller often reaches steady state on its first actuation while the full range controller requires multiple actuation periods to respond to a change in reference.

Identifying operating regions at design time allows us to switch system models at runtime, improving the effectiveness of static controllers.

\begin{figure}[]
\centering
        \subfloat[Power for full frequency range.]{
            \includegraphics[width=0.48\textwidth]{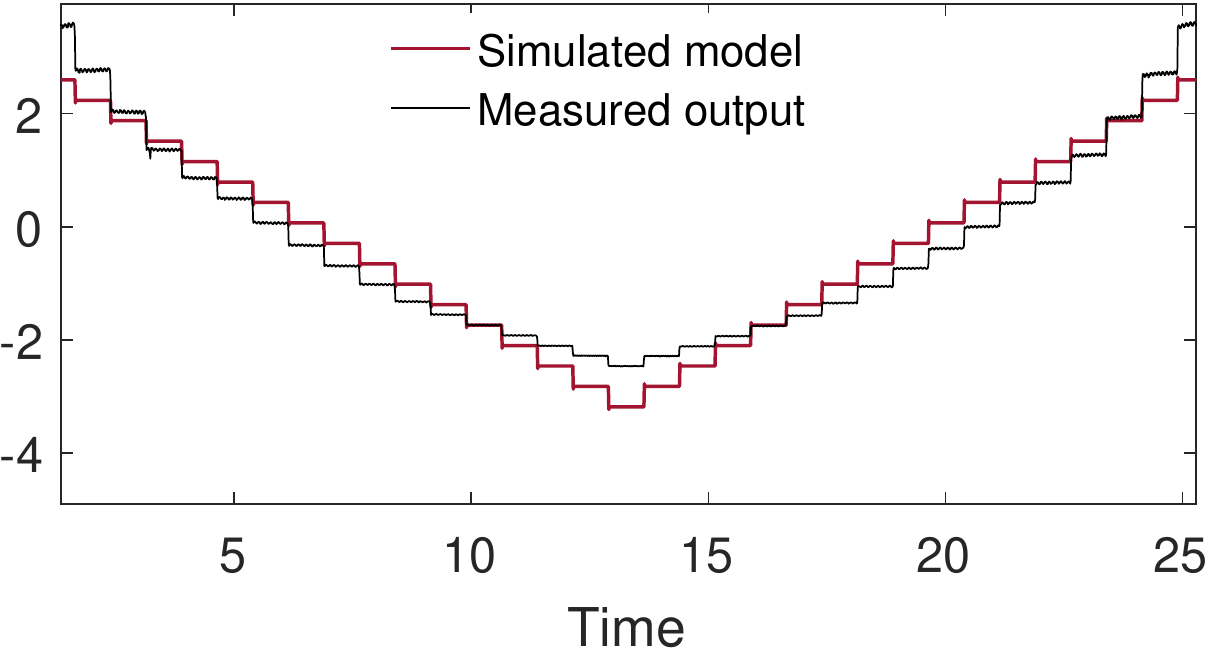}
            \label{fig:motiv:all-1800}
        }%
        \subfloat[Power for 200--800 MHz.]{
                \includegraphics[width=0.48\textwidth]{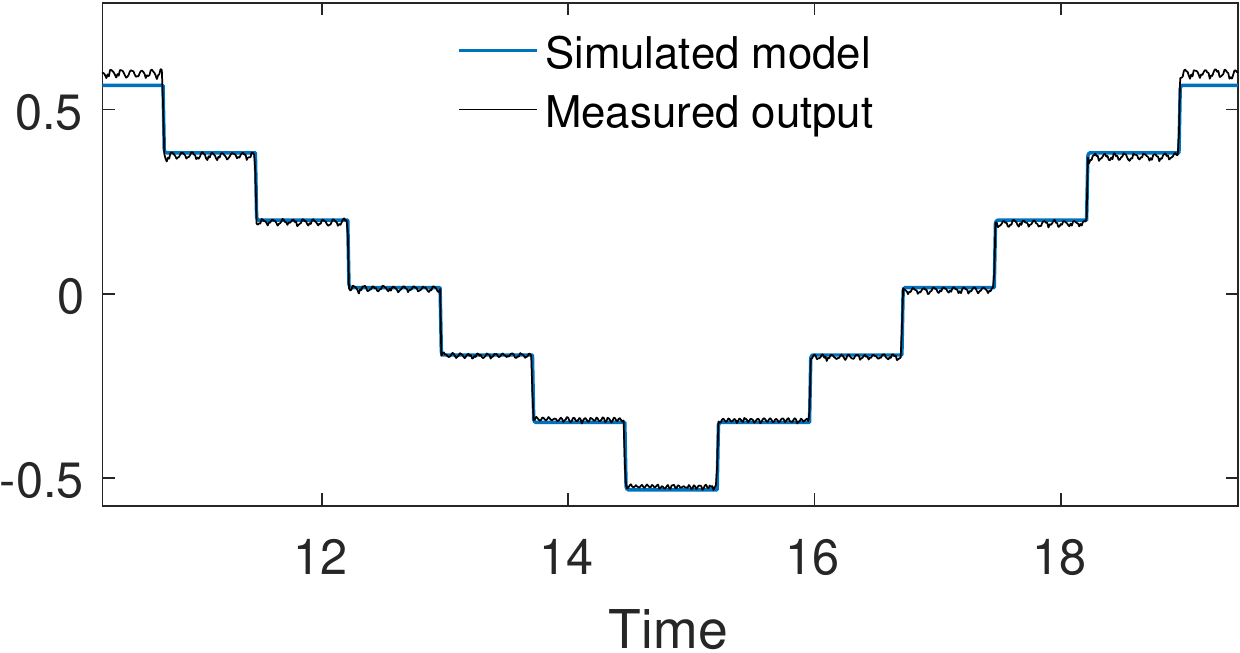}
                \label{fig:motiv:L3}
        }%
        
        \caption{Modeled and observed behavior of nonlinear full-range system (a) vs. linear operating region (b) (from \cite{Donyanavard2018b}).}
        \label{fig:motiv:SI}
\end{figure}

\subsection{Case Study: Designing a Gain Scheduled Controller (GSC) for Power Management}
\label{sec:design}
As a demonstrative case study, we target the ODROID-XU3 platform \cite{Hardkernel2016} which contains an ARM big.LITTLE based Exynos 5422 Octa-core SoC that has heterogeneous multi-processing (HMP) cores.
The Exynos platform contains an HMP with two 4-core clusters: the \emph{big} cluster provides high-performance out-of-order cores, while the \emph{little} cluster provides low-power in-order cores.
For the purpose of our study, we disable the little cluster (due to its linear behavior) and use only the big cores to emulate a uniform nonlinear CMP\footnote{
We refer to this as the Exynos CMP or CMP throughout.
}.

\subsubsection{Defining and Modeling Linear Subsystems}

Selecting the control input and measured output of a DVFS controller is straightforward. 
Frequency is the knob available to the user in software, and power is the metric of interest.
On our Exynos CMP, the operating frequency of cores is set at the cluster level, and power sensors measure power at the cluster level.
A SISO controller is a natural solution, with the entire CMP composing the system under control.

For system identification we generate test waveforms from applications and use statistical black-box methods based on System Identification Theory \cite{Ljung,IMTC} for isolating the deterministic and stochastic components of the system to build the model.

\begin{table}
\sidecaption
  \centering
  \input{tab/a15_v-f_pairs.tex}
  \caption{VF Pairs for ARM A15 in Exynos 5422.}
  \label{tab:vf-pairs}
\end{table}

Figure~\ref{fig:motiv:all-1800} shows a comparison of a simulated model output vs. the measured output over the entire frequency range of our CMP.
It is evident that there are ranges for which the estimated behavior differs from that of the actual system behavior.
We know that voltage has a nonlinear effect on dynamic power ($P=CV^2f$). 
The nonlinear relationship between frequency and voltage pairs through the range of operating frequencies amplifies this effect (Table~\ref{tab:vf-pairs}). 
Table~\ref{tab:vf-pairs} lists all valid VF pairs for the CMP, in which there are only four different voltage levels.
Figure~\ref{fig:motiv:L3} shows the measured vs. modeled output when the system is defined by a single operating region grouped by frequencies that operate at the same voltage level.

\subsubsection{Generating Linear Controllers}
We generate a PI controller separately for each operating region using the system models and MATLAB's Control System toolbox.
This is a straightforward process for a simple off-the-shelf PI controller.

\begin{figure}[]
\centering
        \resizebox{0.8\textwidth}{!}{\input{fig/gsc_block_diagram.tex}}
        \caption{Block diagram of GSC.}
        \label{fig:design:controller}
\end{figure}
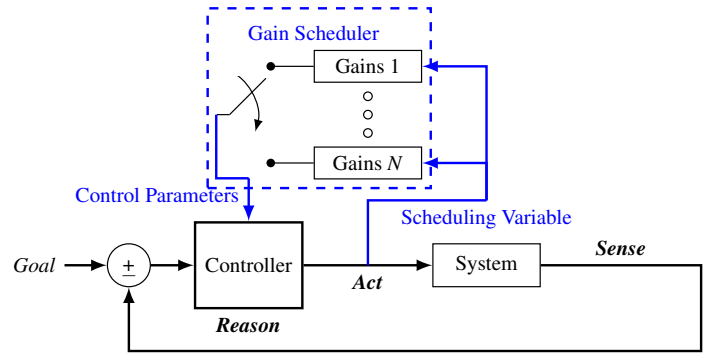

In the next step, the designed controller is evaluated against disturbance and uncertainties in order to ensure it remains stable at a defined confidence level.  
Unaccounted elements, modeling limitations, and environmental effects are estimated as model uncertainty in order to check the disturbance rejection of the controller.
In our case, we can confirm our controller is robust enough to reject the disturbance from workload variation. 

Each controller we design for an operating region is defined by its control parameters $K_P$ and $K_I$ which are stored (in memory) in the gain scheduler (Figure~\ref{fig:design:controller}).
In the gain scheduler, we incorporate logic to determine which gains to provide the controller when invoked.

\subsubsection{Implementing Gain Scheduling}
The gain scheduler enables us to adapt to nonlinear behavior (Figure~\ref{fig:design:controller}) by combining multiple linear controllers.
It stores predefined controller gains and is responsible for providing the most appropriate gains based on the operating region in which the system currently resides each time the controller is invoked.

\begin{algorithm}
  \algsetup{linenosize=\scriptsize}
  \footnotesize
\caption{Gain Scheduler Implementation}
\label{code:gain_sched}
\textbf{Input:} $f$: frequency, scheduling variable;

\textbf{Outputs:} $K_{P_n}$, $K_{I_n}$, $\mathit{offset}_n$: updated controller parameters;

\textbf{Variables:} $\mathit{ref}_{prev}$, $\mathit{ref}_{next}$: power reference values for previous and next control periods;

\textbf{Constants:} $Region[N]$: operating regions, defined by mutually exclusive range of frequencies; $K_P[N]$, $K_I[N]$, $\mathit{offset}[N]$: stored controller parameters for each operating region; $K_{P_G}$, $K_{I_G}$, $\mathit{offset}_G$: controller parameters for full-range linear controller;

\begin{algorithmic}[1]
  \IF{$\mathit{ref}_{next} != ref_{prev}$} \label{code:gain_sched:if1}
    \STATE{$K_{P_n} = K_{P_G}$}
    \STATE{$K_{I_n} = K_{I_G}$}
    \STATE{$\mathit{offset}_n = \mathit{offset}_G$}
    \STATE{$return$}
  \ELSE \label{code:gain_sched:else1}
      \FOR{$i=1$ \TO $N$} \label{code:gain_sched:for}
        \IF{$Region[i].contains(f)$} 
          \STATE{$K_{P_n} = K_P[i]$}
          \STATE{$K_{I_n} = K_I[i]$}
          \STATE{$\mathit{offset}_n = \mathit{offset}[i]$}
          \STATE{$return$}
        \ENDIF
      \ENDFOR \label{code:gain_sched:endfor}
    \ENDIF
\end{algorithmic}
\end{algorithm}

The scheduling variable is the variable used to define operating regions.
For our controller, the scheduling variable is frequency as it is simpler to implement in software and has a direct VF mapping (Table~\ref{tab:vf-pairs}). 
Our gain scheduler implements lightweight logic that determines the set of gains based on the system's operating frequency (scheduling variable).
Algorithm~\ref{code:gain_sched} shows the logic implemented in our gain scheduler with $N$ operating regions where $f$ is the scheduling variable and $K_P$ and $K_I$ are the controller parameters.
In addition to the $K_P$ and $K_I$ controller parameters, there is also an $\mathit{offset}$. 
The $\mathit{offset}$ is the mean actuation value for the operating region, and is necessary for providing the control input for the next control period.
Algorithm~\ref{code:gain_sched} accounts for the transitions between operating regions (lines~\ref{code:gain_sched:if1}-\ref{code:gain_sched:else1}) by applying a full-range linear controller. 
This method is utilized as the sets of gains for a particular operating region perform poorly outside of that region.

\subsubsection{Experiments}
\label{sec:experiments}

Our goal is to evaluate our nonlinear GSC with respect to the state-of-the-art linear controller in terms of both theoretical and observed ability to track power goals on a CMP.
Our evaluation is done using the Exynos CMP running Ubuntu Linux.\footnote{
Ubuntu 16.04.2 LTS and Linux kernel 3.10.105
}
We consider a typical mobile scenario in which one or more multi-threaded applications execute concurrently across the CMP.

\paragraph{Controllers}
We designed two DVFS controllers for power management of the CMP:
1) \textbf{a linear controller} that estimates the transfer function similarly to~\cite{hoffmann2011dynamic,mishra2010coordinated};
and our proposed 2) \textbf{GSC}.
The GSC contains three operating regions (Table~\ref{tab:results:saso}).
We combine the two smallest adjacent Regions, 1 and 2 (Table~\ref{tab:vf-pairs}), to create Controller 2.1.
Controllers are provided a single power reference for the whole system. 
The control input is frequency, and the measured output is power, applied to the entire CMP.

The controller is implemented as a Linux userspace process that executes in parallel with the applications.
Power is calculated using the on-board current and voltage sensors present on the ODROID board.
Power measurements and controller invocation are performed periodically every $50ms$.

\paragraph{Workloads}
We developed a custom micro-benchmark used for system identification.
The micro-benchmark consists of a sequence of independent multiply-accumulate operations yielding varied instruction-level parallelism.
This allows us to model a wide range of behavior in system outputs given changes in the controllable inputs.
We test our controllers using three PARSEC benchmarks: \texttt{bodytrack}, \texttt{streamcluster}, and \texttt{x264}.
For each case, we execute one multithreaded application instance of the benchmark with four threads, resulting in a fully-loaded CMP. 
We empirically select three references that we alternate between during execution.
$\mathit{ref}_1$ is 3.5W, the highest reference and a reasonable power envelope for a mobile SoC. This represents a high-performance mode that maximizes performance under a power budget.
$\mathit{ref}_2$ is 0.5W, the lowest reference and represents a reduced budget in response to a thermal event.
$\mathit{ref}_3$ is 1.5W, a middling reference that could represent the result of an optimizer that maximizes energy efficiency.
These references are not necessarily trackable for all workloads, but should span at least three different operating regions for each workload.
For each case, the applications run for a total of $65s$. 
After the first $5s$ (warm-up period) the controllers are set to $\mathit{ref}_1$ for $20s$, then changed to $\mathit{ref}_2$ for $20s$, and to $\mathit{ref}_3$ for the remaining $20s$.

\subsubsection{Controller Design Evaluation}

\begin{table}[]
  \centering
  \input{tab/dvfs_siso_accuracy.tex}
  \caption{Accuracy of the full- (Ctrl 1) and sub-range (Ctrl 2.x) controllers.}
  \label{tab:results:saso}
\end{table}

We used a first-order system, with a target crossover frequency of 0.32.
This resulted in a simple controller providing the fastest settling time with no overshoot.
Models are generated with a stability focus and uncertainty guardbands of 30\%. 

All systems are stable according to Robust Stability Analysis.
By design all overshoot values are 0.
The settling times of Controllers 2.2 and 2.3 are comparably low at 5 control periods.
Controller 2.1 (the most nonlinear operating region) and Controller 1 are slightly higher at 8-9 control periods.
The ideal controllers are all very similar in terms of stability, settling time, and overshoot.
The primary difference between them is in terms of accuracy.
Controllers 2.1-2.3 achieve an order of magnitude better accuracy than Controller 1 (Table~\ref{tab:results:saso}).
This means that the region controllers are equally as responsive as the full-range model in achieving a target value while achieving the value more accurately. 

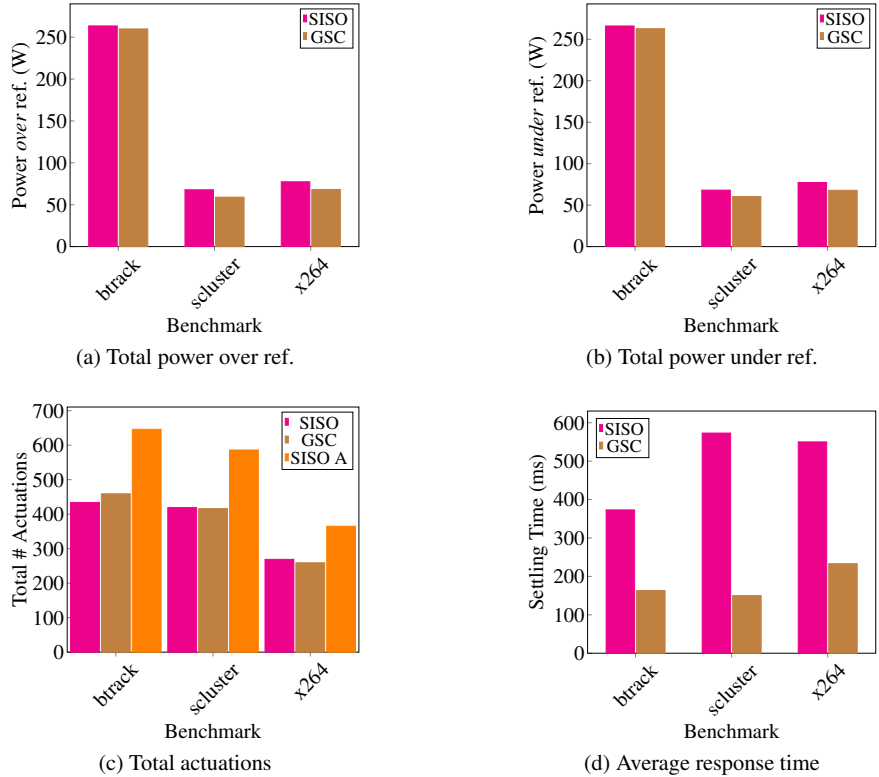
\begin{figure}[]
\centering
        \subfloat[Total power over ref.]{
                \resizebox{0.4\textwidth}{!}{\input{fig/results/gsc-tot_os.tex}}
                \label{fig:results:bar:os}
        }\hfill%
        \subfloat[Total power under ref.]{
                \resizebox{0.4\textwidth}{!}{\input{fig/results/gsc-tot_us.tex}}
                \label{fig:results:bar:us}
        }%

        \subfloat[Total actuations]{
                \resizebox{0.4\textwidth}{!}{\input{fig/results/gsc-act.tex}}
                \label{fig:results:bar:act}
        }\hfill%
        \subfloat[Average response time]{
                \resizebox{0.4\textwidth}{!}{\input{fig/results/gsc-st.tex}}
                \label{fig:results:bar:st}
        }%
        \caption{Comparison of GSC with Controller 1 (from \cite{Donyanavard2018b}).}
        \label{fig:results:bar}
\end{figure}

\subsubsection{Controller Implementation Evaluation}
We now evaluate the effectiveness of our nonlinear control approach implemented in software on the Exynos CMP for multithreaded mobile workloads.
Traditional SASO control analysis gives us a way to compare the controllers in theory, but the system-level effects of those metrics are not directly relatable.
Therefore, we will compare the runtime behavior of the software controllers using a slightly modified set of metrics: power over target, power under target, number of actuations, and response time.
These metrics are shown in Figure~\ref{fig:results:bar}.

The \emph{power over target} is the total amount of measured power exceeding the reference power throughout execution (Fig.~\ref{fig:results:bar:os}). 
This is the area under the output and above the reference.
It represents the amount of power wasted due to inaccuracy, and can also represent unsafe execution above a power cap.
Our GSC is able to achieve \textbf{12\%} less power over target than the linear controller for \texttt{x264} and \texttt{streamcluster}.
\texttt{bodytrack} is the most dynamic workload and results in the noisiest power output. In this case the GSC only improves the power over target by 1\% compared to the linear controller.

The \emph{power under target} is the total amount of measured power falling short of the reference power throughout execution (Fig.~\ref{fig:results:bar:us}). 
This is the area under the reference and above the output.
A lower value translates to improved performance (i.e. lower is better).
Similarly to the power over target, our GSC is able to reduce power under target by \textbf{12\%} for \texttt{x264} and \texttt{streamcluster}, and 1\% for \texttt{bodytrack}.

The \emph{number of actuations} is simply a count of how many times the frequency changes throughout execution, and is a measure of overhead (Fig.~\ref{fig:results:bar:act}).
The GSC's actuation overhead is lower than the linear controller for
\texttt{bodytrack}, 
\texttt{streamcluster}, and \texttt{x264} by 8\%, 1\%, and 4\% respectively.
This is expected, as the controller's resistance to actuation is related to the crossover frequency specified at design time.
For the same crossover frequency, the GSC benefits are primarily in the accuracy (power over/under target) and response (settling) time.
To illustrate this tradeoff, we performed the same experiments for a full-range linear controller with a target crossover frequency of 0.8 (Controller 1b).
We arrived at this value empirically: Controller 1b achieves comparable accuracy to the GSC.
However, GSC reduces the actuation overhead by \textbf{29\%} for all workloads compared to Controller 1b.

The \emph{response time} is the average settling time when the target power changes, indicating the controller's ability to respond quickly to changes (Fig.~\ref{fig:results:bar:st}).
Figure~\ref{fig:results:bar:st} shows the average response time for each workload for both controllers.
The GSC is able to improve the response time over Controller 1 by more than \textbf{50\%} in each case.
The GSC's overall average response time is $182ms$, which is less than 4 control periods.

The implementation overhead of the GSC w.r.t. the linear controller is negligible: it requires a single execution of Algorithm~\ref{code:gain_sched} upon each invocation, and storage for a $K_P$, $K_I$, and $\mathit{offset}$ value for each operating region.
Although workload disturbance plays a significant role in determining the magnitude in imporovement of a nonlinear GSC over a state-of-the-art linear controller, a clear trend exists, and these advantages would increase with the modeled system's degree of nonlinearity.

\subsection{Summary}
Self-models are the core components of self-awareness. 
In computer systems, system dynamics can be complex. 
When utilizing a self-model at runtime for reflection, models must be simple and sufficiently accurate. 
The more accurate the self-model, the more effective the decisions made by a resource manager can be toward achieving a given goal. 
We propose a simple way to improve the accuracy of self-models for resource managers employing classical controllers: gain scheduling. 
Gain scheduled control generates multiple controllers based on optimized fixed models for different operating regions of the system, and can deploy the most accurate control at runtime based on the system state.
This is an improvement over using a single controller based on a single fixed model with minimal overhead. 
In our case study, the gain scheduled controller more effectively provides dynamic power management of a single-core processor when compared to a single fixed SISO controller.

Using a static model for resource management may not be sufficient in complex mobile systems: system dynamics may change between applications or devices, and fixed models may not remain accurate over time.
In the future, we plan to address such scenarios by identifying and continuously updating models during runtime based on observation, instead of identifying multiple fixed models at design-time to swap out at runtime.


\section{Self-adaptivity}

Self-adaptivity is the ability of a system to adjust to changes in goals due to external stimuli.
For example, if a system experiences a thermal event during a computational sprint and enters an unsafe state, a self-adaptive manager will have the ability to modify the goal from maximizing performance to minimizing temperature.

\subsection{Motivation and Background}

To address self-optimization, we examined a relatively simply use-case in which we deployed a resource manager responsible for controlling a unicore with only a single input and single output.
However, modern computer systems incorporate up to hundreds of cores, from datacenters to mobile devices.
Modern mobile devices commonly deploy architecturally differentiated cores on a single chip multiprocessor, known as heterogeneous multiprocessors (HMPs).
In the case of mobile devices, systems are tasked with the challenge of balancing application goals with system constraints, e.g., a performance requirement within a power budget.
Resource managers are required to configure the system at runtime to meet the goal.
However, due to workload or operating condition variation, it is possible for goals to change unpredictably at runtime.
In this section, we use self-adaptivity to enable a mobile HMP resource manager to adapt to a changing goal, coordinating and prioritizing multiple objectives.

\subsubsection{Managing Dynamic System-wide Goals}
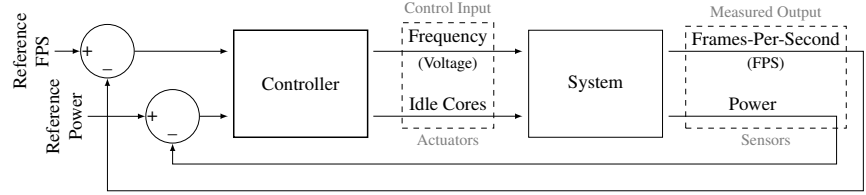
\begin{figure}[]
 \begin{center}
    \resizebox{\textwidth}{!}{\input{fig/MIMO_2x2.tex}}    
\end{center}
	\caption{Basic $2 \times 2$ MIMO for single-core system. Clock frequency and idle cores are used as control inputs. FPS and power are measured outputs that are compared with reference (i.e., target) values.}
	\label{fig:basic_mimo}
\end{figure}

Controllers may behave non-optimally, or even detrimentally, in meeting a shared goal without knowledge of the presence or behavior of seemingly orthogonal controllers~\cite{HDGM,Bitirgen:2008:CMM,Vega:2013:CUD,Ebrahimi:2009:CCM:1669112.1669154,Ebrahimi:2010,Das:2009}.
\label{qr}
Consider the MIMO controller in Figure~\ref{fig:basic_mimo} that controls a single-core system with two control inputs (\textit{u(t)}) and interdependent measured outputs (\textit{y(t)})~\cite{MIMO-16}. 
The controller tracks two objectives (frames per second, or FPS, and power consumption) by controlling two actuators (operating frequency and cache size).
We implement the MIMO using a Linear Quadratic Gaussian (LQG) controller~\cite{Skogestad:2005:MFC} similarly to~\cite{MIMO-16}:
%
%
\begin{eqnarray}
	x(t+1)&=&A \times x(t)+B \times u(t) \label{eq.lqg}\\
	y(t)&=&C \times x(t)+D \times u(t)   \label{eq.lqg2}
\end{eqnarray} 
\noindent where $x$ is the system state, $y$ is the measured output vector, and $u$ is the control input vector.\footnote{We interchangeably use the terms (\textit{measured output} and \textit{sensor}), as well as the  terms (\textit{control input} and \textit{actuator}), as shown in Figure~\ref{fig:basic_mimo}.}

LQG control allows us to specify 1) the relative sensitivity of a system to control inputs, and 2) the relative priority of measured outputs.
This is done using 1) a weighted Tracking Error Cost matrix (\textit{Q}) and 2) a Control Effort Cost matrix (\textit{R}).
The weights are specified during the design of the controller.
While this is convenient for achieving a fixed goal, it can be problematic for goals that change over time (e.g., minimizing power consumption before a predicted thermal emergency).

The controller must choose an appropriate trade-off when we cannot achieve both desirable performance and power concurrently.
Unfortunately, classical MIMOs fix control weights at design time, and thus \emph{cannot} perform \emph{runtime} tradeoffs that require changing output priorities.
Even with constant \emph{reference values}, i.e., desired output values, unpredictable disturbances (e.g., changing workload and operating conditions) may cause the reference values to become unachievable.
It is also plausible for the reference values themselves to change dynamically at runtime with system state and operating conditions (e.g., a thermal event).

Let us now consider a more complex scenario: a multi-threaded application running on Linux, executing on a mobile processor, where the system needs to track both the performance (FPS) and power simultaneously.
Figure~\ref{fig:basic_mimo} shows the $2 \times 2$ MIMO model for this system with operating frequency and the number of active cores as control inputs, and FPS and power as measured outputs.

\begin{figure}[]
        \subfloat[FPS-oriented controller.]{
                \includegraphics[width=0.50\columnwidth]{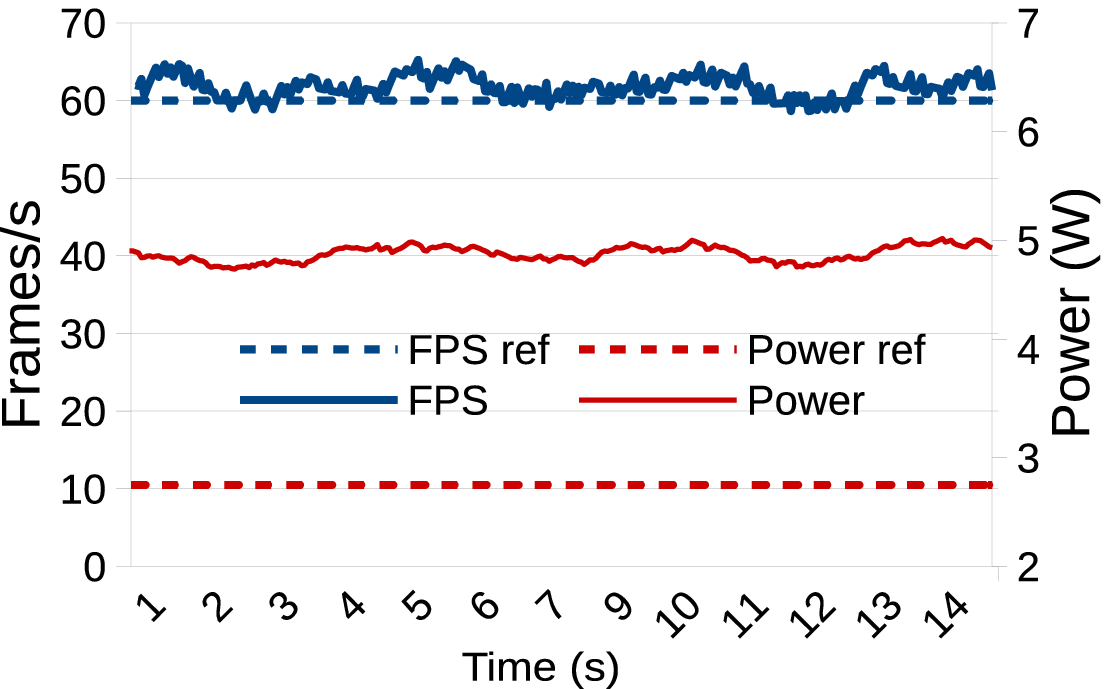}
                \label{fig:gain_scheduling:a}
        }%
        \hfill
        \subfloat[Power-oriented controller.]{
                \includegraphics[width=0.50\columnwidth]{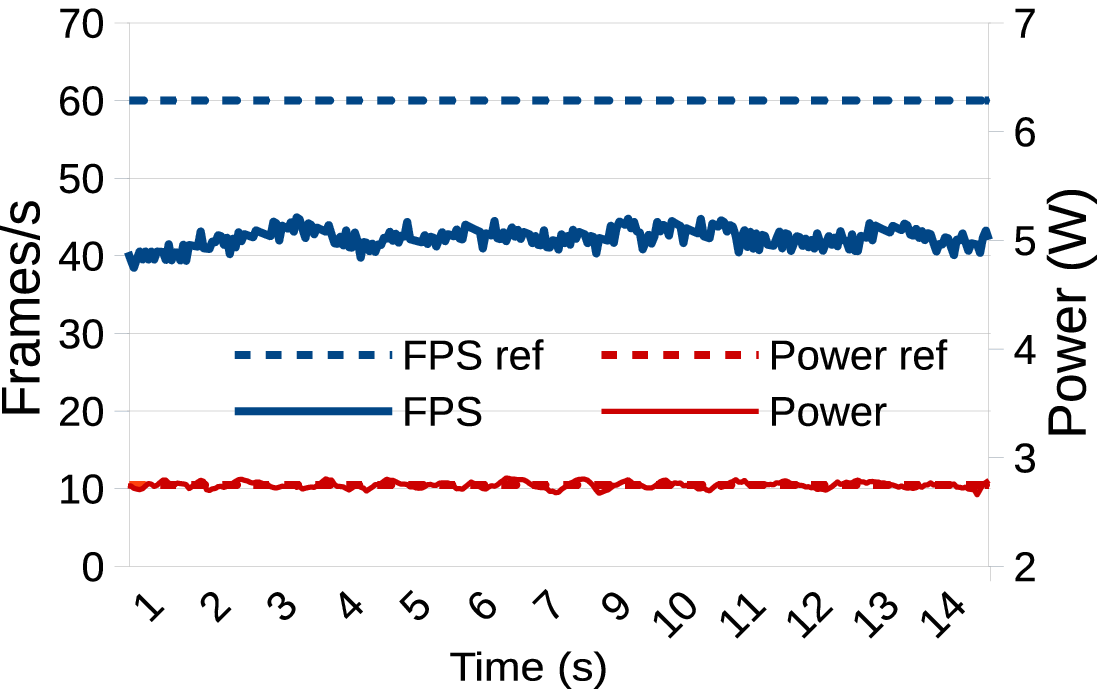}
                \label{fig:gain_scheduling:b}
        }%
        \caption{\texttt{x264} running on a quad-core cluster controlled by $2 \times 2$ MIMOs with different output priorities (from \cite{Rahmani2018}).}
\label{fig:gain_scheduling}
\end{figure}

Both the FPS and power reference values are trackable individually, but not jointly.
We implement and compare two different MIMO controllers in Linux to show the effect of competing objectives.
One controller prioritizes FPS, and the other prioritizes power.
Figure~\ref{fig:gain_scheduling} shows the power and performance (in FPS) achieved by each MIMO controller using typical reference values for a mobile device: 60 FPS and 5 Watts.
The application is \texttt{x264}, and the mobile processor consists of an ARM Cortex-A15 quad-core cluster.
Each MIMO controller is designed with a different \textit{Q} matrix
to prioritize either FPS or power:
Figure~\ref{fig:gain_scheduling:a}'s controller favors FPS over power by a ratio of 30:1 (i.e., only 1\% deviation from the FPS reference is acceptable for a 30\% deviation from the power reference), while Figure~\ref{fig:gain_scheduling:b} uses a ratio of 1:30.
We observe that neither controller is able to manage changing system goals. 
Thus, there is a need for a supervisor to autonomously orchestrate the system while considering the significance of competing objectives, user requirements, and operating conditions.

The use of supervisory control presents at least three additional advantages over conventional controllers.
First, fully-distributed MIMO or SISO controllers \emph{cannot} address system-wide goals such as power capping. 
Second, conventional controllers \emph{cannot} model actuation effects that require system-wide perspective, such as task migration.
Third, classical control theory \emph{cannot} address problems requiring optimization (e.g., minimizing an objective function) alone \cite{Karamanolis:2005:DCC,MIMO-16}.

\subsubsection{Supervisory Control Theory}\label{SCT}

Supervisory control utilizes modular decomposition to mitigate the complexity of control problems, enabling automatic control of many individual controllers or control loops. 
Supervisory control theory (SCT)~\cite{SCT1989} benefits from formal synthesis methods to define principal control properties for \textit{controllability} and \textit{observability}. 
The emphasis on formal methods in addition to \textit{modularity} leads to \textit{hierarchical consistency} and \textit{non-conflicting} properties.


SCT solves complex synthesis problems by breaking them into small-scale sub-problems, known as modular synthesis. 
The results of modular synthesis characterize the conditions under which decomposition is effective.
In particular, results identify whether a valid decomposition exists.
A decomposition is valid if the solutions to sub-problems combine to solve the original problem, and
the resulting composite supervisors are \textit{non-blocking} and \textit{minimally restrictive}.
Decomposition also adds robustness to the design because nonlinearities in the supervisor do \textit{not} directly affect the system dynamics. 

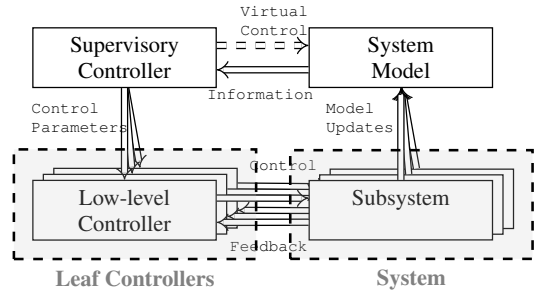
\begin{figure}[]
 \begin{center}
 \resizebox{0.6\textwidth}{!}{\input{fig/sct_overview.tex}}
 \end{center}
	\caption{Supervisory Control structure. Low level control loops are guided by the Supervisory Controller that achieves system-wide goals based on the high-level system model.}
	\label{fig:sct}
\end{figure}

Figure \ref{fig:sct} illustrates how a supervisory control structure can hierarchically manage control loops.
As shown in the figure, supervision is vertically decomposed into tasks performed at different levels of abstraction \cite{THISTLE199625}. 
The supervisory controller is designed to control the high-level \textit{system model}, which represents an abstraction of the system.
The \textit{subsystems} compose the pre-existing \textit{system} that does \textit{not} meet the given specifications without the aid of a controller or a supervisor.
The \textit{information} channel provides information about the updates in the high-level model to the supervisory controller.
Due to the fact that the system model is an abstract model, the controlling channel is an indirect \textit{virtual control} channel. 
In other words, the control decisions of the supervisory controller will be implemented by controlling the \textit{low-level controller(s)} through \textit{control parameters}.
Consequently, the low-level controller(s) can control one or multiple subsystems using the \textit{control} channel and gather information via \textit{feedback}.
The changes in the subsystems can trigger \textit{model updates} in the state of the high-level system model. 
These updates reflect the results of low-level controllers' controlling actions.

The scheme of Figure \ref{fig:sct} describes the division of supervision into high-level management and low-level operational supervision. 
Virtual control exercised via the high-level control channel can be implemented by modifying control parameters to adaptively coordinate the low-level controllers, e.g., by adjusting their objective functions according to the system goal. 
The combination of horizontal and vertical decomposition enables us to not only physical divide the system into subsystems, but also to logically divide the sub-problems in any appropriate way, e.g., due to varying epochs (control invocation period) or scope.
The important requirement of this hierarchical control scheme is control consistency and hierarchical consistency between the high-level model and the low-level system, as defined in the standard Ramadge-Wonham control mechanism~\cite{THISTLE199625}.  
For a detailed description of SCT, we refer the reader to \cite{SCT1989,Safonov1997,Brandin:1991,THISTLE199625}.

\subsubsection{Self-Adaptivity via Supervisory Control}
Supervisory controllers are preferable to \textit{adaptive (self-tuning) controllers} for complex system control due to their ability to integrate \textbf{logic} with \textbf{continuous dynamics}. 
Specifically, supervisory control has two key properties: i) rapid adaptation in response to abrupt changes in management policy 
\cite{HespanhaNov01}, and ii) low computational complexity by computing control parameters for different policies \textbf{offline}.
New policies and their corresponding parameters can be added to the supervisor on demand (e.g., by upgrading the firmware or OS), 
rendering online learning-based self-tuning methods, e.g., least-squares estimation \cite{Astrom}, unnecessary.

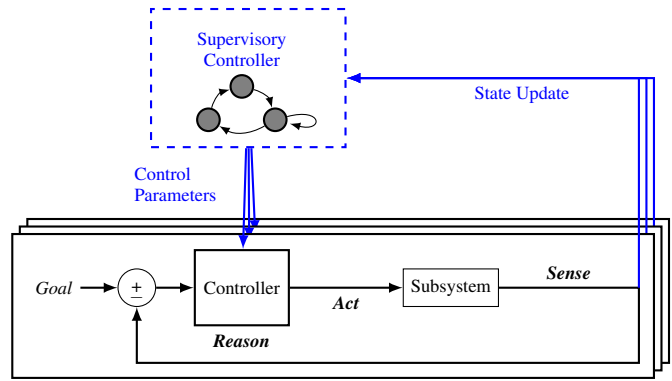
\begin{figure}[b]
 \begin{center}
 \input{fig/sct_gsc.tex}
 \end{center}
	\caption{Self-adaptivity via gain scheduling in SCT.}
    	\label{fig.asbtract-sc2}
\end{figure}

Figure \ref{fig.asbtract-sc2} depicts the two mechanisms that enable SCT-based management via low-level controllers: 
\textbf{gain scheduling} and \textbf{dynamic references}.
Gain scheduling is a nonlinear control technique that uses a set of linear controllers predesigned for different operating regions.
Gain scheduling enables the appropriate linear controller based on runtime observations \cite{GainScheduling}. 
Scheduling is implemented by switching between sets of control parameters, i.e., $A_1$$\rightarrow$$A_2$, $B_1$$\rightarrow$$B_2$, $C_1$$\rightarrow$$C_2$, and $D_1$$\rightarrow$$D_2$ in Equations \ref{eq.lqg} and \ref{eq.lqg2}.
In this case, the \textit{controller gains} are the values of the control parameters $A$, $B$, $C$, and $D$.
Gains are useful to change objectives at runtime in response to abrupt and sudden changes in management policy.
In LQG controllers, this is done by changing priorities of outputs using the \textit{Q} and \textit{R} matrices (Section~\ref{qr}). 
This is what we call the Hierarchical Control structure, in which local controllers solve specified tasks while the higher-level supervisory controller coordinates the global objective function. 
In this structure, the supervisory controller receives information from the plant (e.g., the presence of a thermal emergency) or the user/application (e.g., new QoS reference value), and steers the system towards the desired policy using its design logic and high-level model.
Thanks to its top-level perspective, the supervisor can update reference values for each low-level controller to either optimize for a certain goal (e.g., getting to the optimum energy-efficient point) or manage resource allocation (e.g., allocating power budget to different cores).

\subsection{Case Study: On-chip Resource Management}\label{SPECTR}

In this section, we design and evaluate a supervisor used to implement a hierarchical resource managemer. 
The use-case requires management of QoS under a power budget on a HMP.
The resource manager (\supervisory) consists of a supervisor that guides low-level classical controllers to configure core operating frequency and number of active cores for each core cluster.

\subsubsection{Hierarchical System Architecture}
\label{HSA}

 Figure~\ref{fig.sc4} depicts a high-level view of \supervisory{} for many-core system resource management.
 Either the user or the system software may specify \emph{Variable Goals and Policies}. 
 The \emph{Supervisory Controller} aims to meet system goals by managing the low-level controllers. 
 High-level decisions are made based on the feedback given by the \emph{High-level Plant Model}, which provides an abstraction of the entire system. 
 Various types of \emph{Classic Controllers}, such as PID or state-space controllers, can be used to implement each low-level controller based on the target of each subsystem.
 The flexibility to incorporate any pre-verified off-the-shelf controllers without the need for system-wide verification is essential for the modularity of this approach. 
 The supervisor provides parameters such as output references or gain values to each low-level controller during runtime according to the system policy.
 Low-level controller subsystems update the high-level model to maintain global system state, and potentially trigger the supervisory controller to take action. 
 The high-level model can be designed in various fashions (e.g., rule-based or estimator-based~\cite{Safonov1997}\cite{HespanhaNov01}\cite{Morse1997}) to track the system state and provide the supervisor with guidelines. 
 We illustrate the steps for designing a supervisory controller using the following experimental case study in which SCT is deployed on a real 
 HMP platform, and we then outline the entire design flow from modeling of the high-level plant to generating the supervisory controller.

\begin{figure}[b]
 \centering
 \resizebox{\textwidth}{!}{\input{fig/SPECTR_overview}}
	\caption{\supervisory{} overview.}
	\label{fig.sc4}
\end{figure}

\subsubsection{\supervisory{} Resource Manager}
\label{bigLittleCaseStudy}

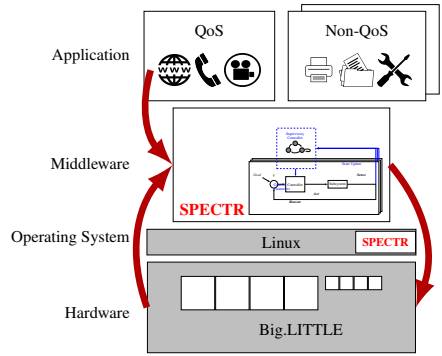
\begin{figure}[h]
 \begin{center}
 \resizebox{0.5\textwidth}{!}{\input{fig/spectr_platform.tex}}
 \end{center}
 \vspace{-10pt}
	\caption{\supervisory{} implementation on the Exynos HMP with two heterogeneous quad-core clusters. Representing a typical mobile scenario with a single foreground application running concurrently with many background applications.}
	\label{fig:case_study-hmp}
\end{figure}

Figure~\ref{fig:case_study-hmp} shows an overview of our experimental setup.
We target the Exynos platform \cite{Hardkernel2016}, which contains an HMP
with two quad-core clusters: the \textbf{Big} core cluster provides high-performance out-of-order cores, while the \textbf{Little} core cluster provides low-power in-order cores.
Memory is shared across all cores, so application threads can transparently execute on any core in any cluster.
We consider a typical mobile scenario in which a single foreground application (the \emph{QoS application}) is running concurrently with many background applications (the \emph{Non-QoS applications}).
This mimics a typical mobile use-case in which gaming or media processing is performed in the foreground in conjunction with background email or social media syncs.

\textbf{The system goals} are twofold: i) meet the QoS requirement of the foreground application while minimizing its energy consumption; and ii) ensure the total system power always remains below the 
Thermal Design Power (TDP). 

The \textbf{subsystems} are the two heterogeneous quad-core (\emph{Big} and \emph{Little}) clusters.
Each cluster has two actuators: one actuator to set the operating frequency (\textit{F\textsubscript{next}}) and associated voltage of the cluster; and one to set the number of active cores (\textit{AC\textsubscript{next}}) on the cluster. 
We measure the power consumption (\textit{P\textsubscript{curr}}) of each cluster, and simultaneously monitor the QoS performance (\textit{QoS\textsubscript{curr}}) of the designated application to compare it to the required QoS (\textit{QoS\textsubscript{ref}}).\footnote{
The Exynos platform provides only per-cluster power sensors and DVFS; hence our use of cluster-level sensors and actuators.
}

Supervisory control commands guide the \textbf{low-level MIMO controllers} in Figure~\ref{fig:case_study-hmp} to determine the number of active cores and the core operating frequency within each cluster.

\textbf{Supervisory control} minimizes the system-wide power consumption while maintaining QoS. 
In our scenario, the QoS application runs only on the Big cluster, and the supervisor determines whether and how to adjust the cluster's power budget based on QoS measurements. 

\textbf{Gain scheduling} is used to switch the priority objective of the low-level controllers. 
We define two sets of gains for this case-study: 
1) \emph{QoS-based} 
gains are tuned to ensure that the QoS application can meet the performance reference value, and
2) \emph{Power-based} gains are tuned to limit the power consumption while possibly sacrificing some performance if the system is exceeding the power budget threshold.

\subsubsection{Experimental Evaluation}
\label{case_study}

We compare \supervisory{} with three alternative resource managers.
The first two managers use two uncoordinated 2$\times$2 MIMOs, one for each cluster: \textit{\multiPow} uses power-oriented gains, and \textit{\multiPerf} uses performance-oriented gains.
These fixed MIMO controllers act as representatives of a state-of-the-art solution, as presented in \cite{MIMO-16}, one prioritizing power and the other prioritizing performance.
The third manager consists of a single full-system controller (\textit{\fs}): a system-wide 4$\times$2 MIMO with individual control inputs for each cluster.
\textit{\fs} uses power-oriented gains and its measured outputs are chip power and QoS.
This single system-wide MIMO acts as a representative for \cite{Zhang:2016:MPU:2872362.2872375}, maximizing performance under a power cap.

We analyze an execution scenario that consists of three different phases of execution:
\begin{enumerate}
\item \emph{Safe Phase}: In this phase, only the QoS application executes (with an achievable QoS reference within the TDP). 
The goal is to meet QoS and minimize power consumption.
\item \emph{Emergency Phase}: In this phase, the QoS reference remains the same as that in the Safe Phase while the power envelope is reduced (emulating a thermal emergency). 
The goal is to adapt to the change in reference power while maintaining QoS (if possible).
\item \emph{Workload Disturbance Phase}: In this phase, the power envelope returns to TDP and background tasks are added (to induce interference from other tasks). 
The goal is to meet the QoS reference value without exceeding the power envelope.
\end{enumerate}
This execution scenario with three different phases allows us to evaluate how \supervisory{} compares with state-of-the-art resource managers when facing workload variation and system-wide changes in state (e.g., thermal emergency) and goals.

\paragraph{Evaluated resource manager configurations}
We generate stable low-level controllers for each resource manager using the Matlab System Identification Toolbox~\cite{MatlabsysIden}.\footnote{
We generate the models with a stability focus. 
All systems are stable according to Robust Stability Analysis.
We use Uncertainty Guardbands of 50\% for QoS and 30\% for power, as in \cite{MIMO-16}.
}
We use the Control Effort Cost matrix (\textit{R}) to prioritize changing clock frequency over number of cores at a ratio of 2:1, as frequency is a finer-grained and lower-overhead actuator than core count.
We generate training data by executing an in-house microbenchmark and varying control inputs in the format of a staircase test (i.e., a sine wave), both with single-input variation and all-input variation.
The micro-benchmark consists of a sequence of independent multiply-accumulate operations performed over both sequentially and randomly accessed memory locations, thus yielding various levels of instruction-level and memory-level parallelism.
The range of exercised behavior resembles or exceeds the variation we expect to see in typical mobile workloads, which is the target application domain of our case studies.

\paragraph{Experimental setup}
We perform our evaluations on the ARM big.LITTLE~\cite{ARM2013c} based Exynos SoC (ODROID-XU3 board~\cite{Hardkernel2016}) as described in our case study (Figure \ref{fig:case_study-hmp}).
We implement a Linux userspace daemon process that invokes the low-level controllers every $50ms$. 
When evaluating \supervisory, the daemon invokes the supervisor every $100ms$.
We use ARM's Performance Monitor Unit (PMU) and per-cluster power sensors for the performance and power measurements required by the resource managers.
The userspace daemon also implements the Heartbeats API~\cite{Hoffmann2013a} monitor to measure QoS.
By periodically issuing \emph{heartbeats}, the application informs the system about its current performance.
The user provides a performance reference value using the Heartbeats API. 

To evaluate the resource managers, we use the following benchmarks from the PARSEC benchmark suite~\cite{bienia11benchmarking} as QoS applications (i.e., the applications that issue heartbeats to the controller):
\texttt{x264}, \texttt{bodytrack}, \texttt{canneal}, and 
\texttt{streamcluster}.
The selected applications consist of the most CPU-bound along with the most cache-bound PARSEC benchmarks, providing varied responses to change in resource allocation.
Speedups from $3.2X$ (\texttt{streamcluster}) to $4.5X$ (\texttt{x264}) are observed with the maximum resource allocation values compared to the minimum.
We also use one of four machine-learning workloads as our QoS application: \texttt{k-means}, \texttt{KNN}, \texttt{least squares}, and \texttt{linear regression}.
These four workloads provide a wide range of data-intensive use cases.
For all experiments, each QoS application uses four threads.
The background (non-QoS) tasks used in the third execution phase are single-threaded microbenchmarks, and have no runtime restrictions, i.e., the Linux scheduler can freely migrate them between and within clusters.

\subsubsection{Effectiveness of Self-Adaptivity through Supervision}

\begin{figure}[]
        \subfloat[\multiPow{} FPS]{
                \includegraphics[width=0.5\columnwidth]{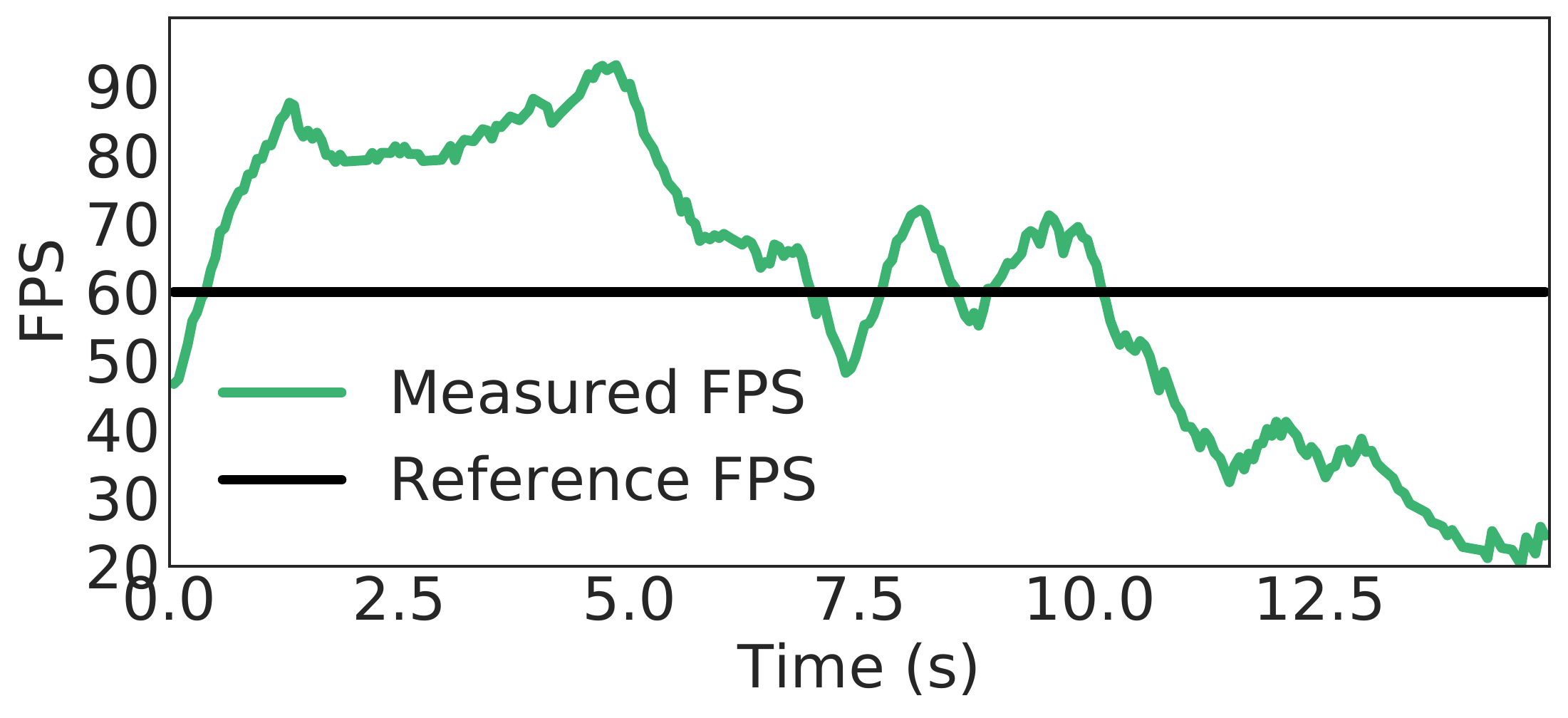}
                \label{fig:results:mm-power:qos}
        }%
        \subfloat[\multiPow{} Power]{
                \includegraphics[width=0.5\columnwidth]{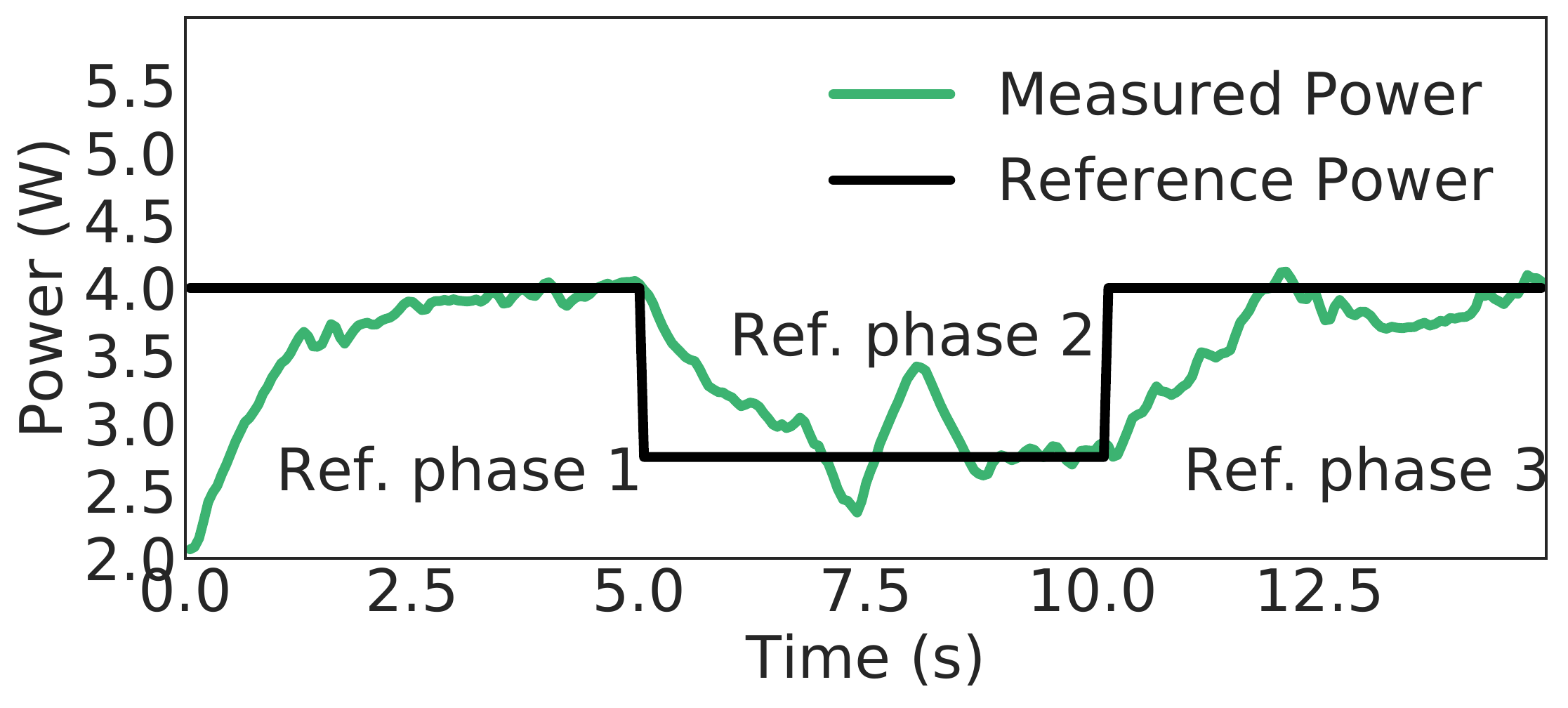}
                \label{fig:results:mm-power:power}
        }%
        
        \subfloat[\multiPerf{} FPS]{
                \includegraphics[width=0.5\columnwidth]{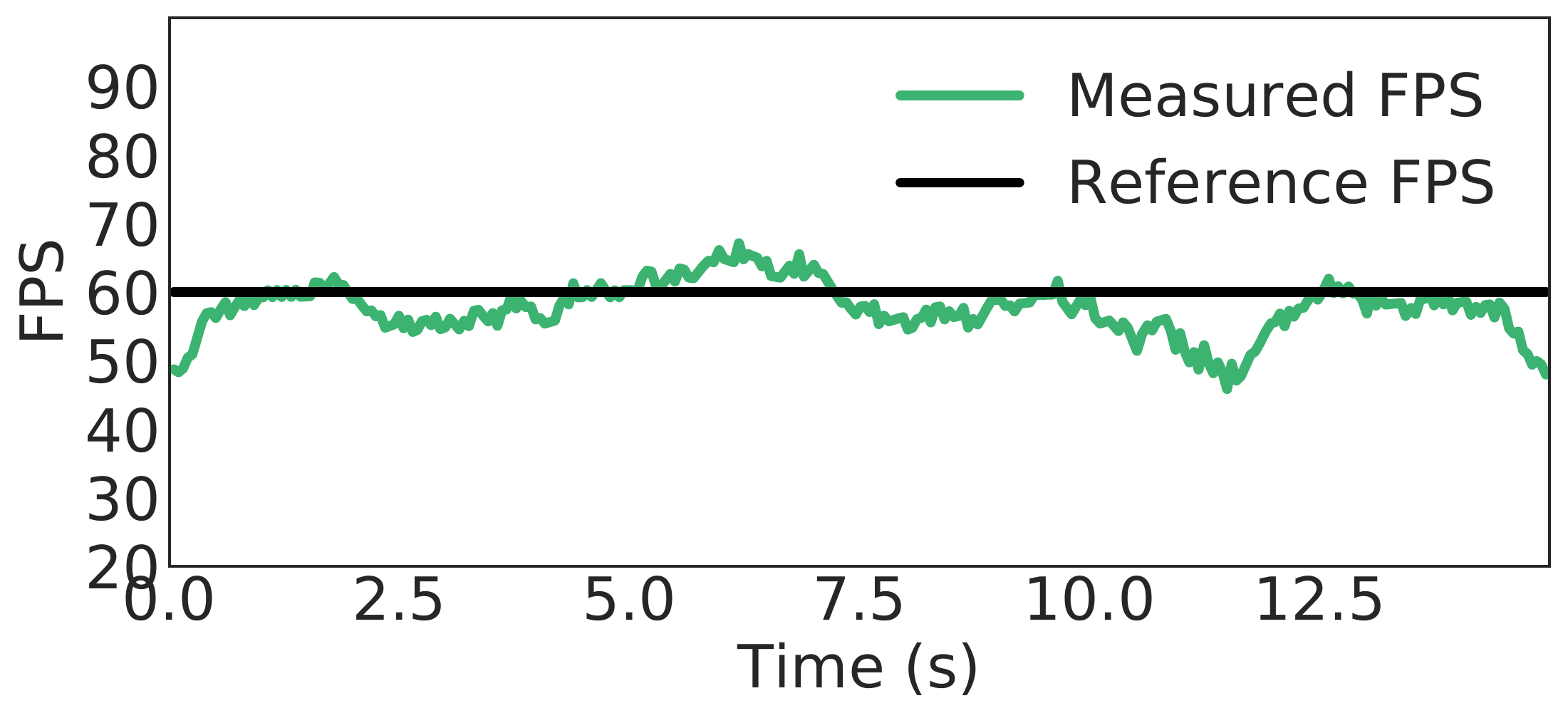}
                \label{fig:results:mm-perf:qos}
        }%
        \subfloat[\multiPerf{} Power]{
                \includegraphics[width=0.5\columnwidth]{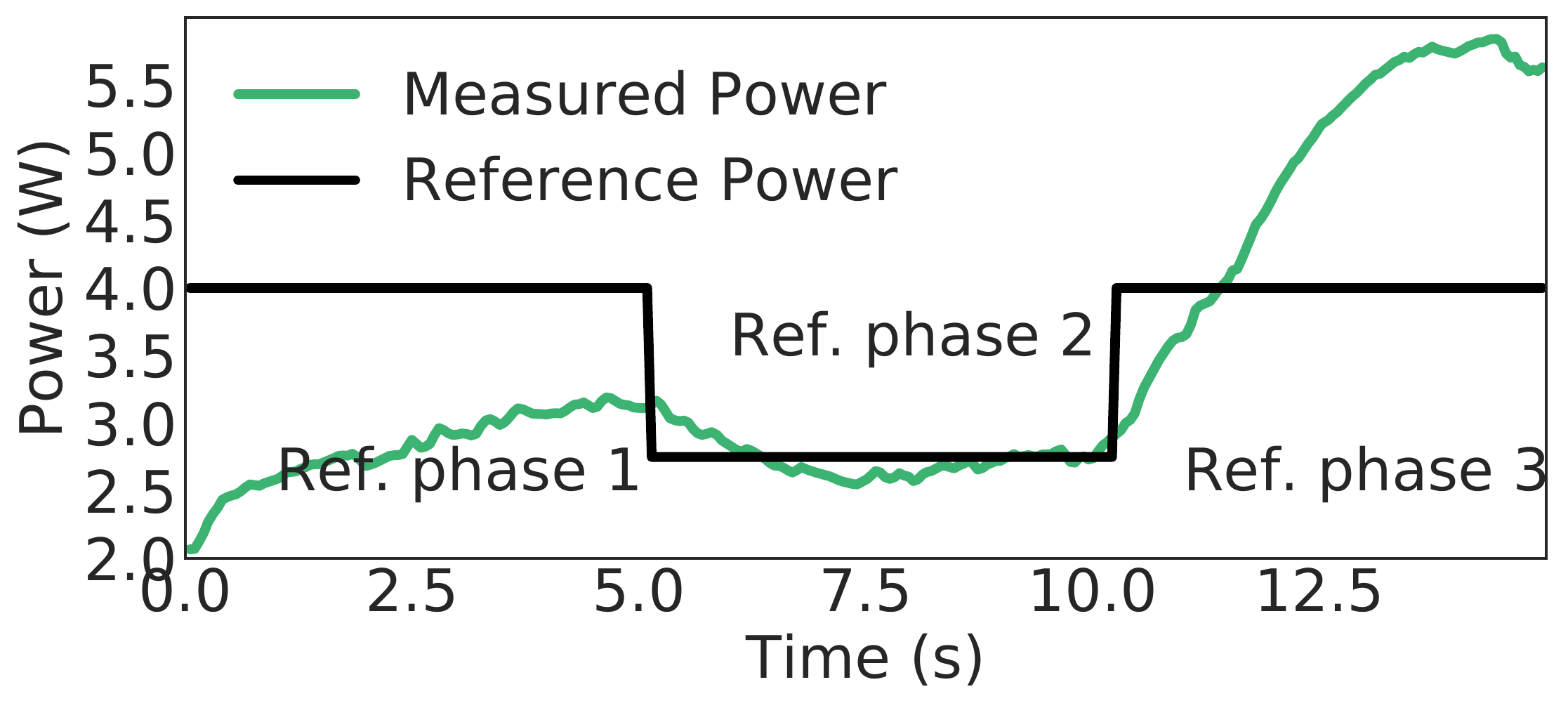}
                \label{fig:results:mm-perf:power}
        }%
        
        \subfloat[\fs{} FPS]{
                \includegraphics[width=0.5\columnwidth]{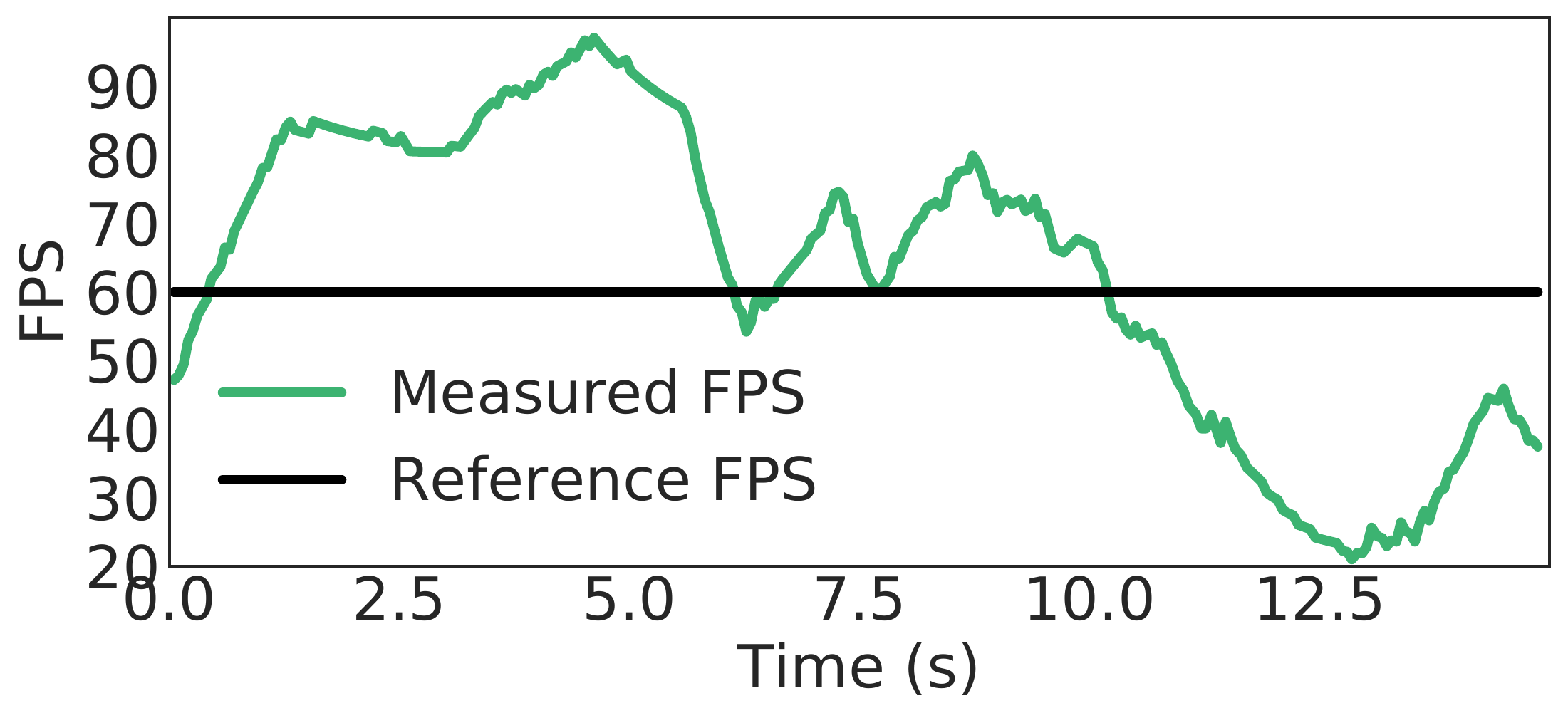}
                \label{fig:results:fs:qos}
        }%
        \subfloat[\fs{} Power]{
                \includegraphics[width=0.5\columnwidth]{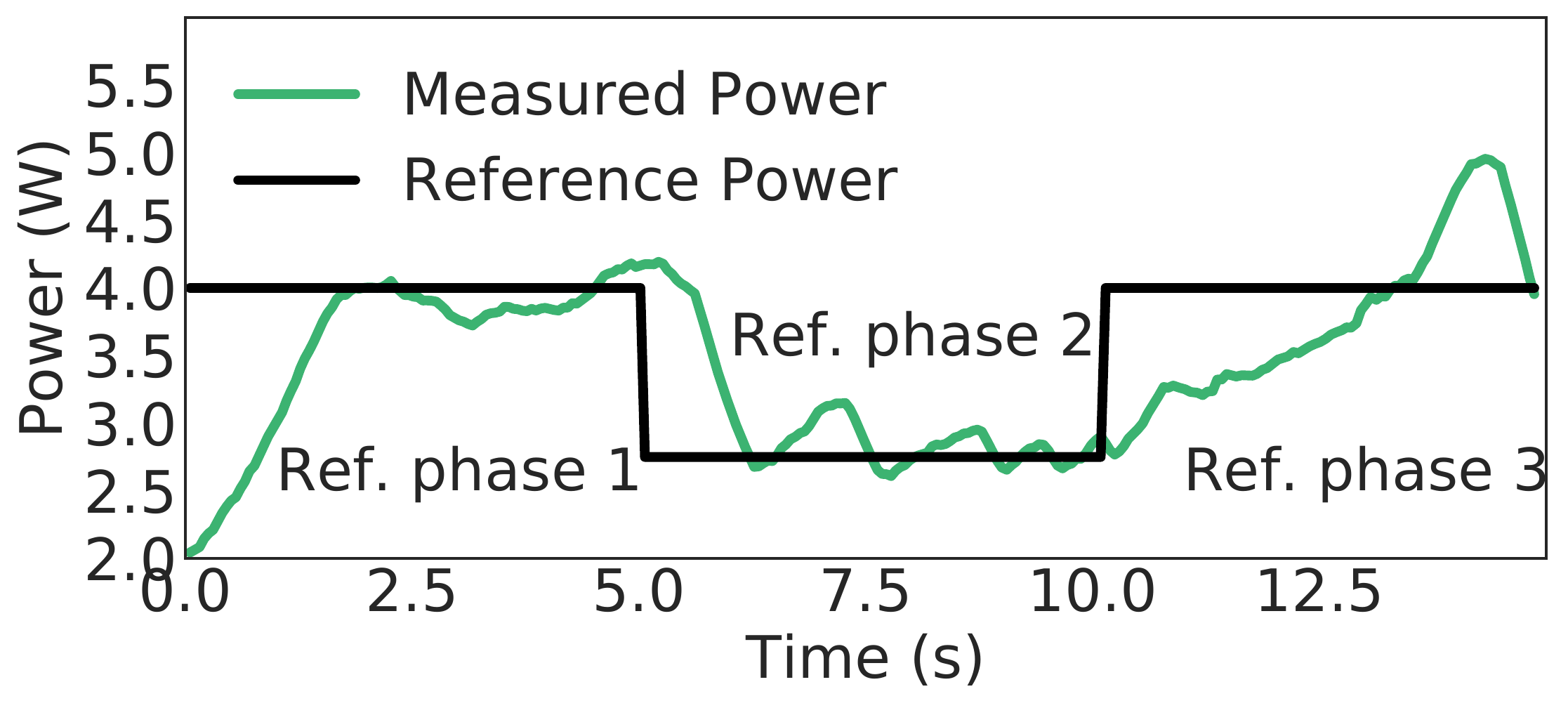}
                \label{fig:results:fs:power}
        }%
        
        \subfloat[\supervisory{} FPS]{
                \includegraphics[width=0.5\columnwidth]{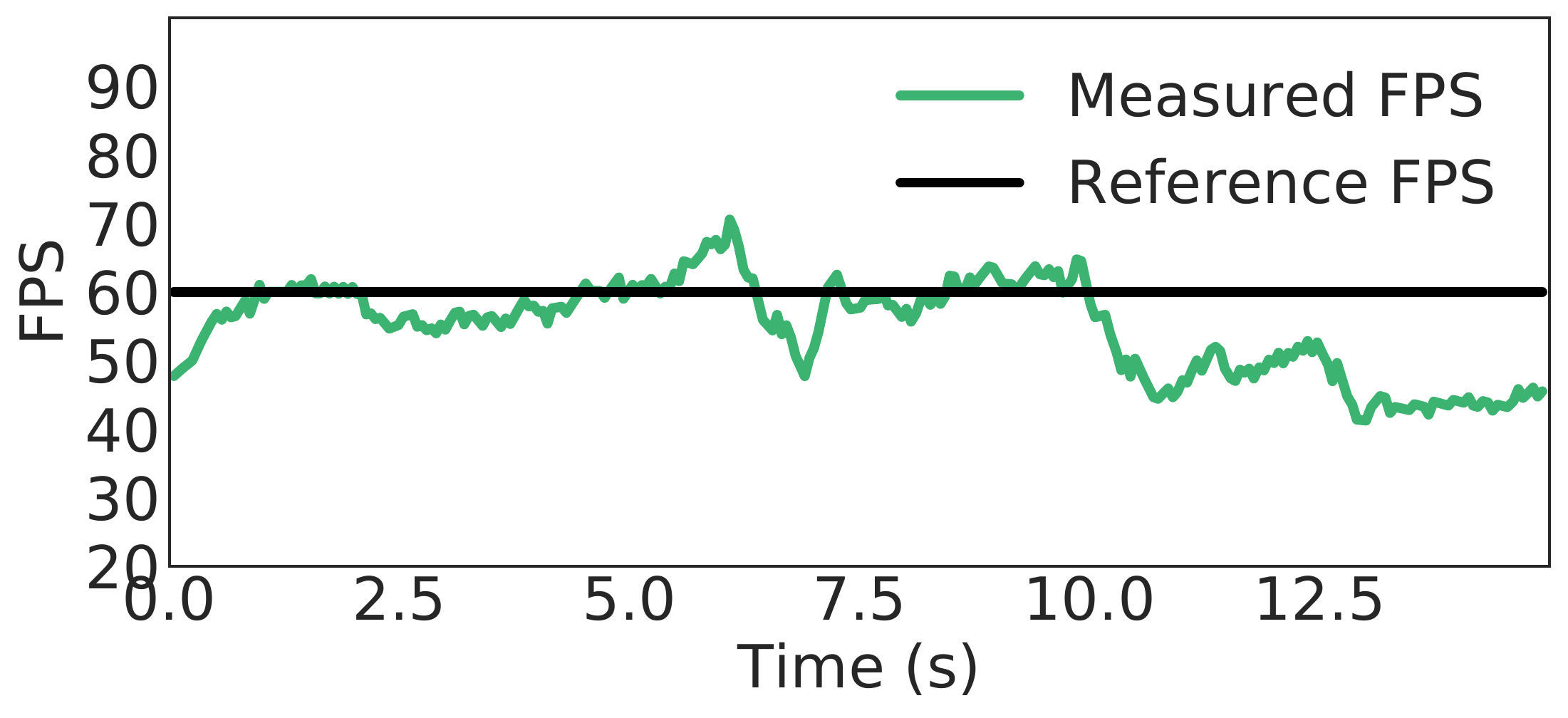}
                \label{fig:results:time:spectr:qos}
        }%
        \subfloat[\supervisory{} Power]{
                \includegraphics[width=0.5\columnwidth]{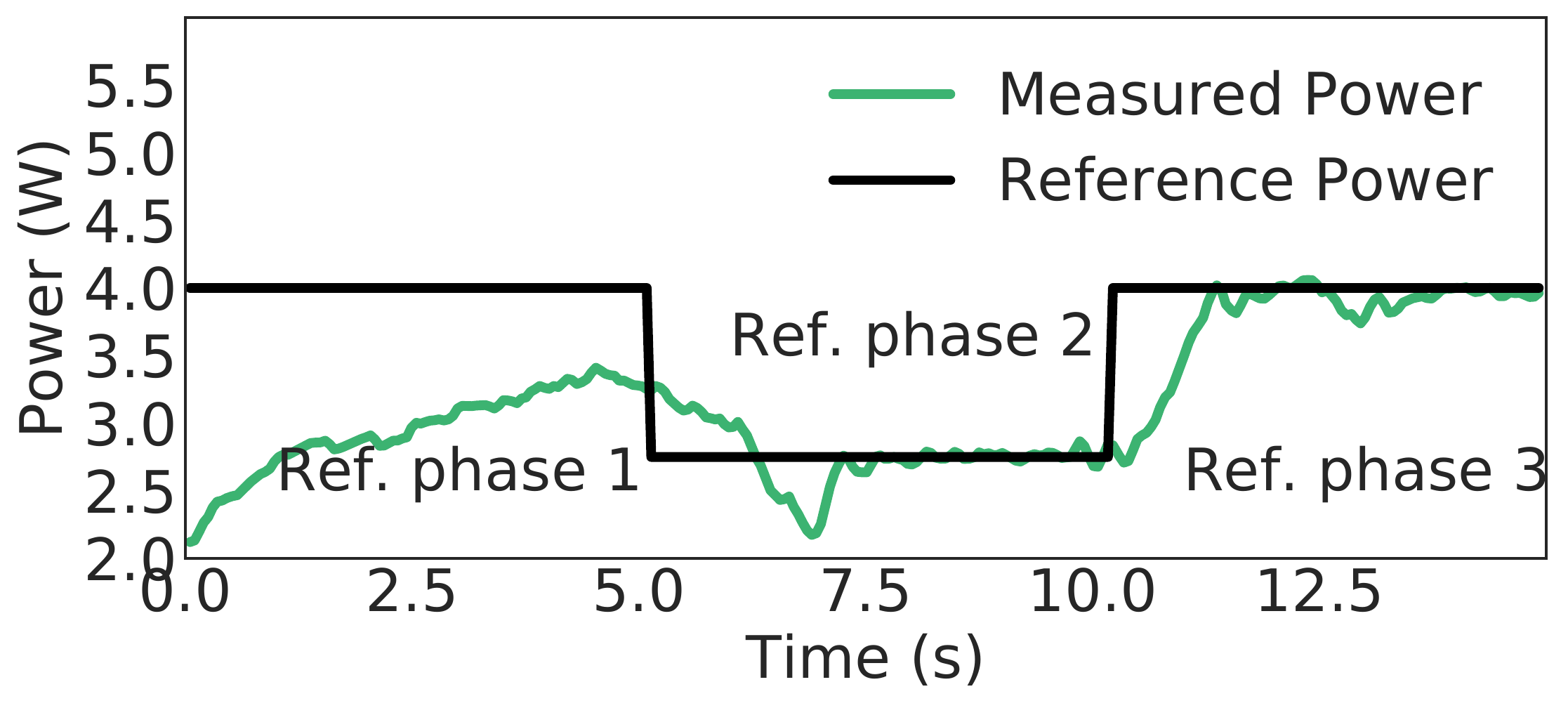}
                \label{fig:results:time:spectr:power}
        }%
        \caption{Measured FPS and Power of all four resource managers for three Phases of 5 seconds each, for the \texttt{x264} benchmark (from \cite{Rahmani2018}).}
        \label{fig:results:time}
\end{figure}

We focus our discussion on the \texttt{x264} benchmark results.
Other results are summarized at the end of this section.
We use heartbeats to measure the frames per second (FPS) as our QoS metric.
Figure~\ref{fig:results:time} shows the measured FPS and power for \texttt{x264} with respect to their reference values over the course of execution for all of the resource management controllers.

\begin{figure*}[hbt]
        \centering
        \subfloat[QoS steady-state error in Phase 1.]{
                \includegraphics[width=0.46\textwidth]{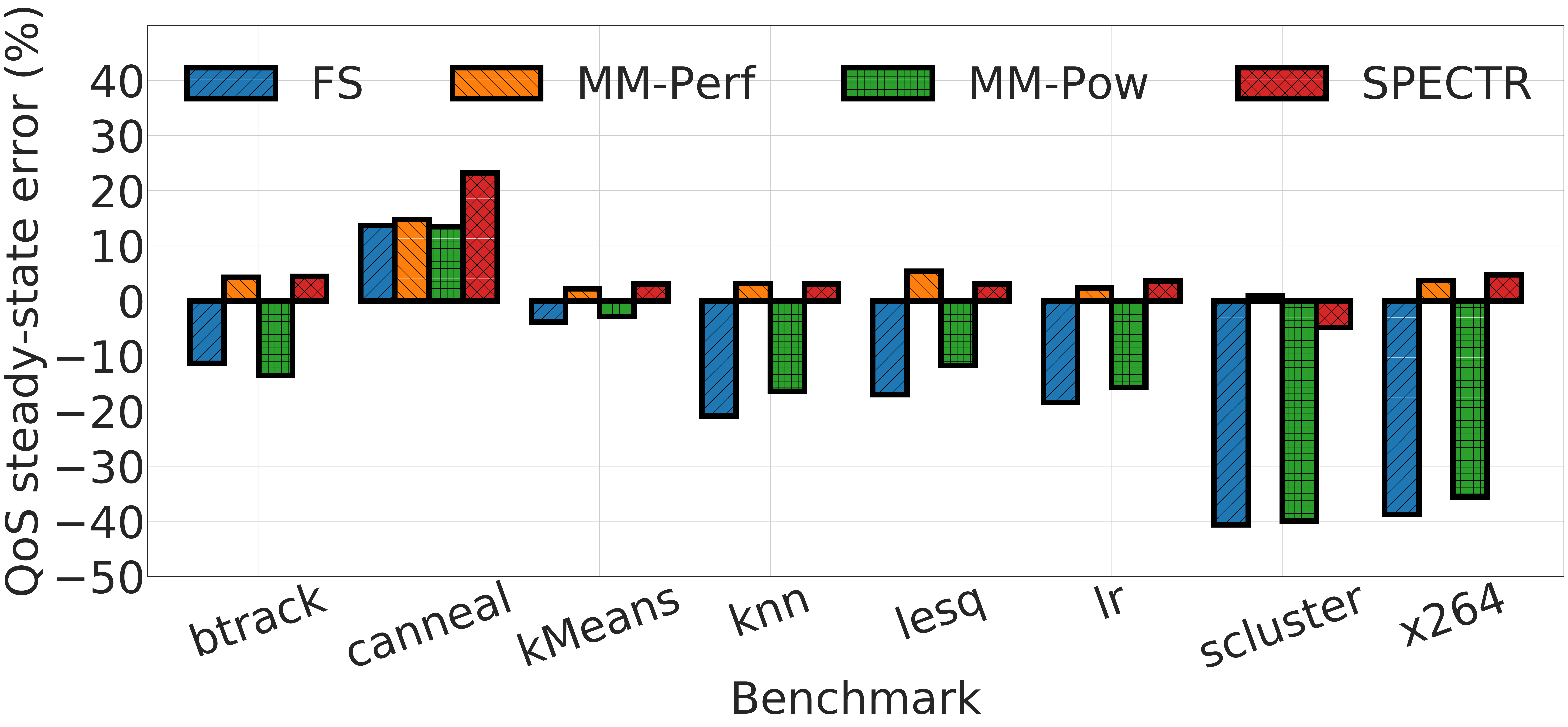}
                \label{fig:results:ss:ph1:qos}
        }\hfill
        \subfloat[Power steady-state error in Phase 1.]{
                \includegraphics[width=0.46\textwidth]{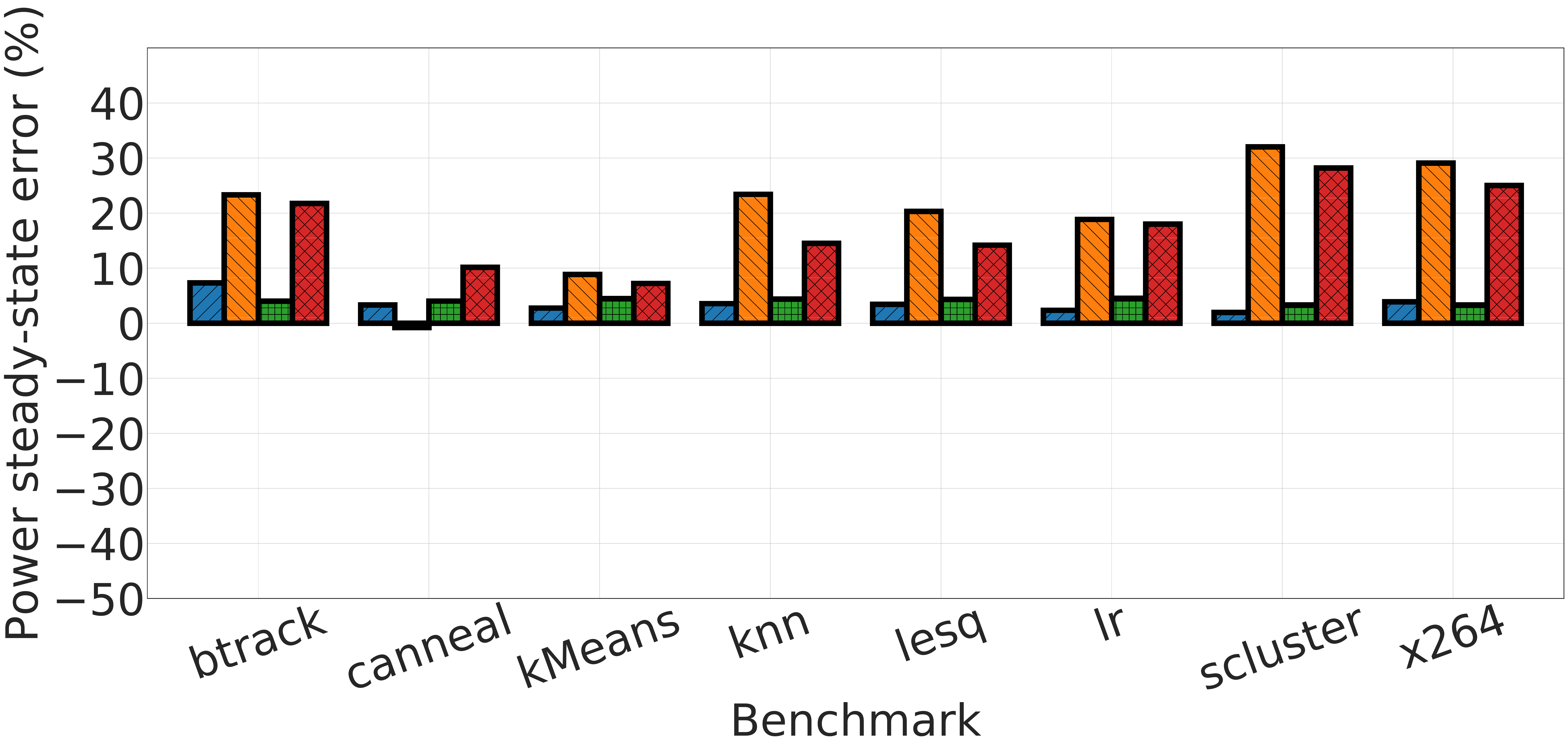}
                \label{fig:results:ss:ph1:pow}
        }%
        
        \subfloat[QoS steady-state error in Phase 2.]{
                \includegraphics[width=0.46\textwidth]{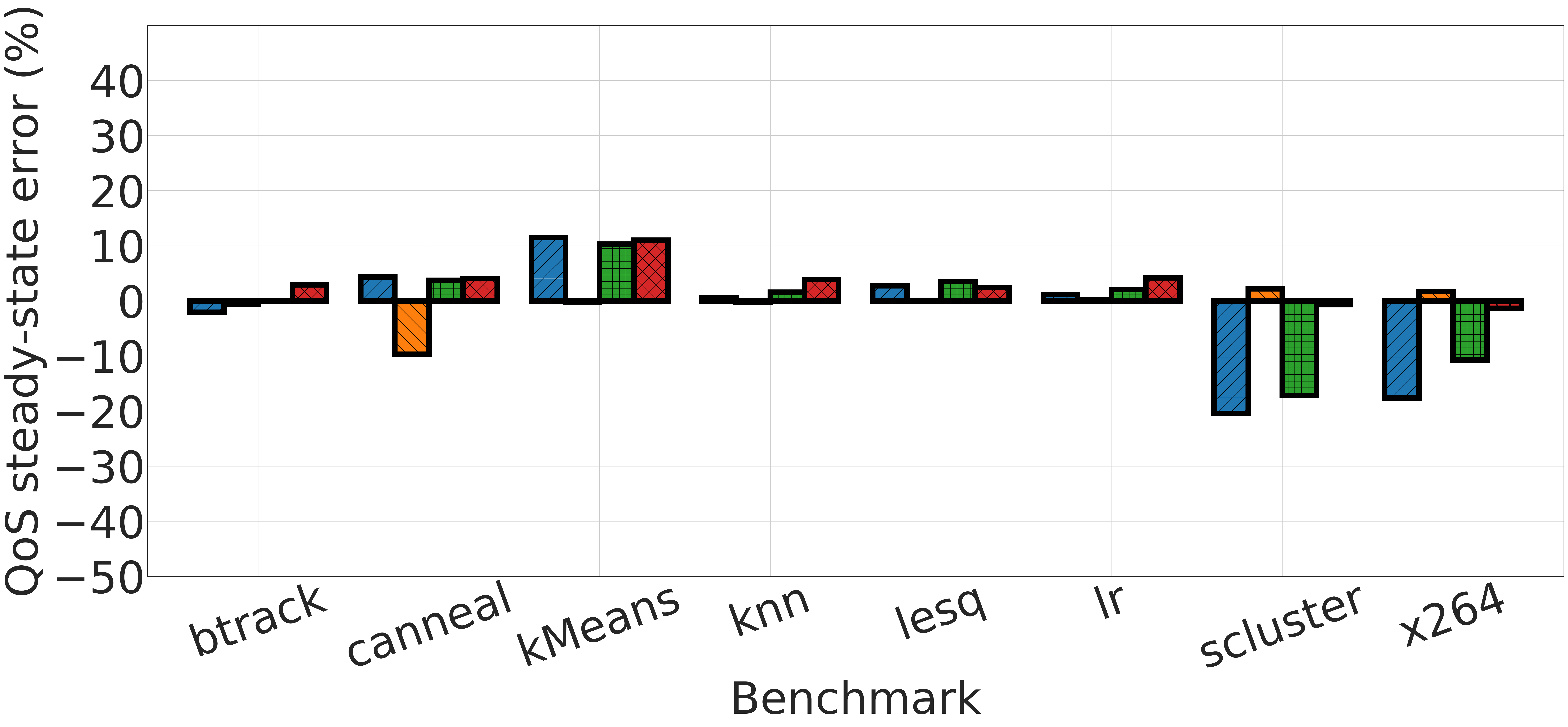}
                \label{fig:results:ss:ph2:qos}
        }\hfill
        \subfloat[Power steady-state error in Phase 2.]{
                \includegraphics[width=0.46\textwidth]{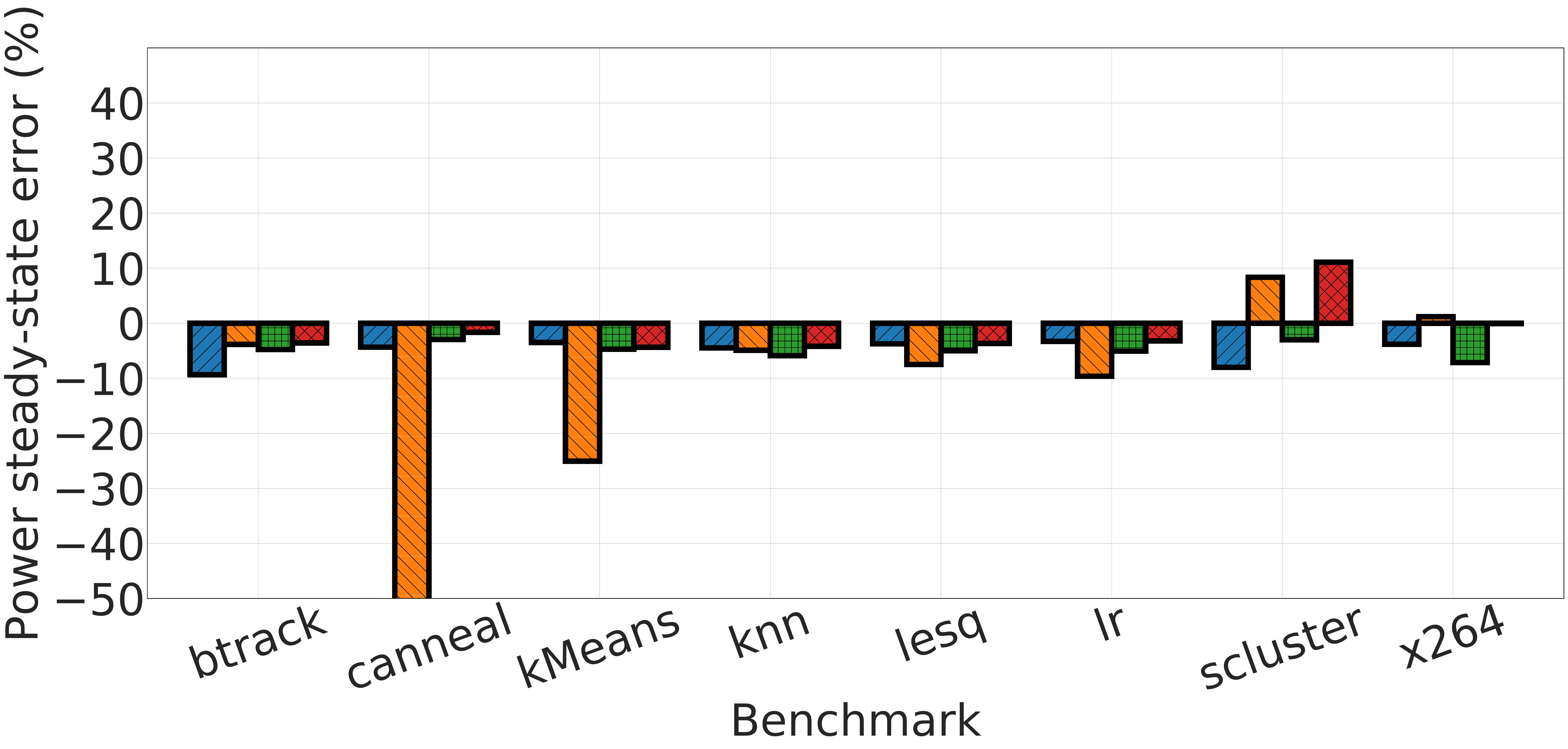}
                \label{fig:results:ss:ph2:pow}
        }%

        \subfloat[QoS steady-state error in Phase 3.]{
                \includegraphics[width=0.46\textwidth]{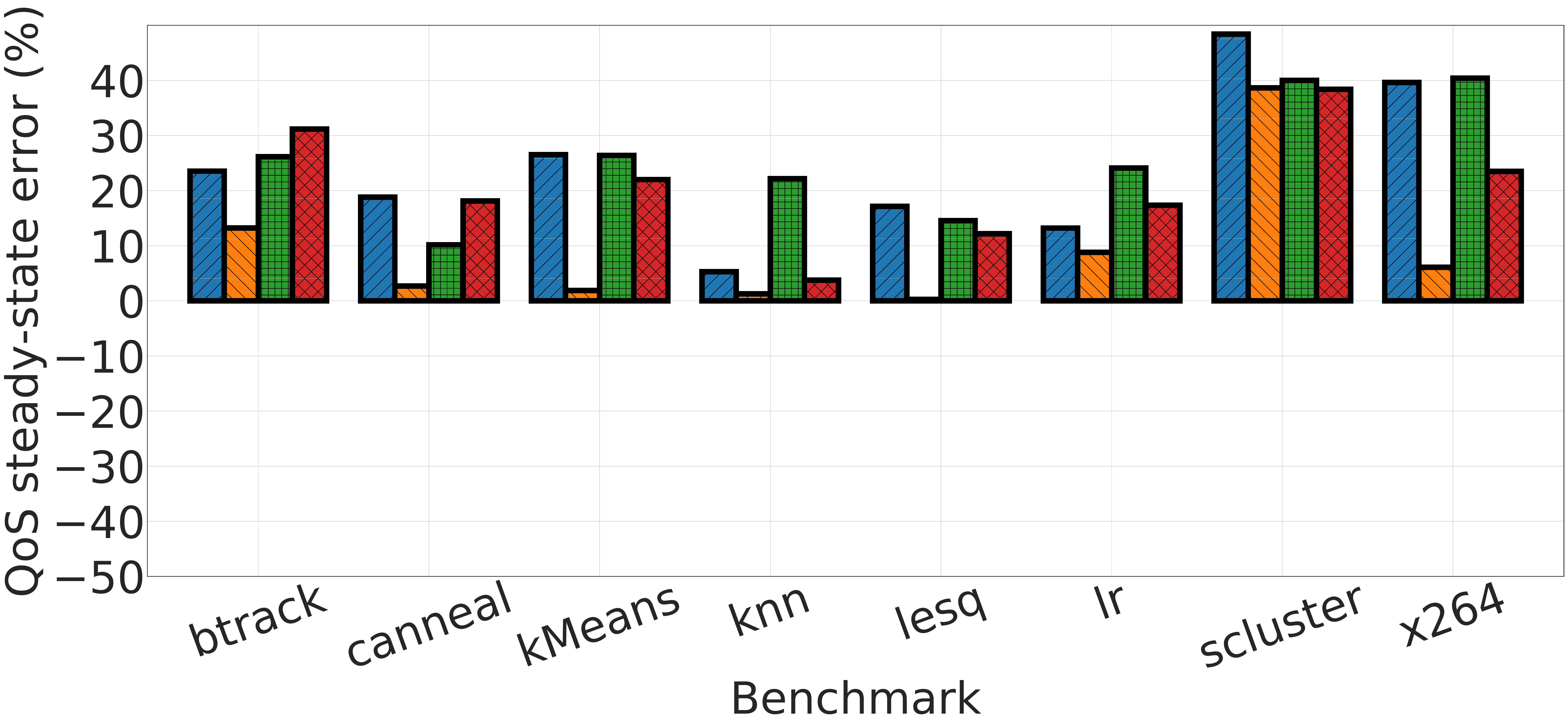}
                \label{fig:results:ss:ph3:qos}
        }\hfill
        \subfloat[Power steady-state error in Phase 3.]{
                \includegraphics[width=0.46\textwidth]{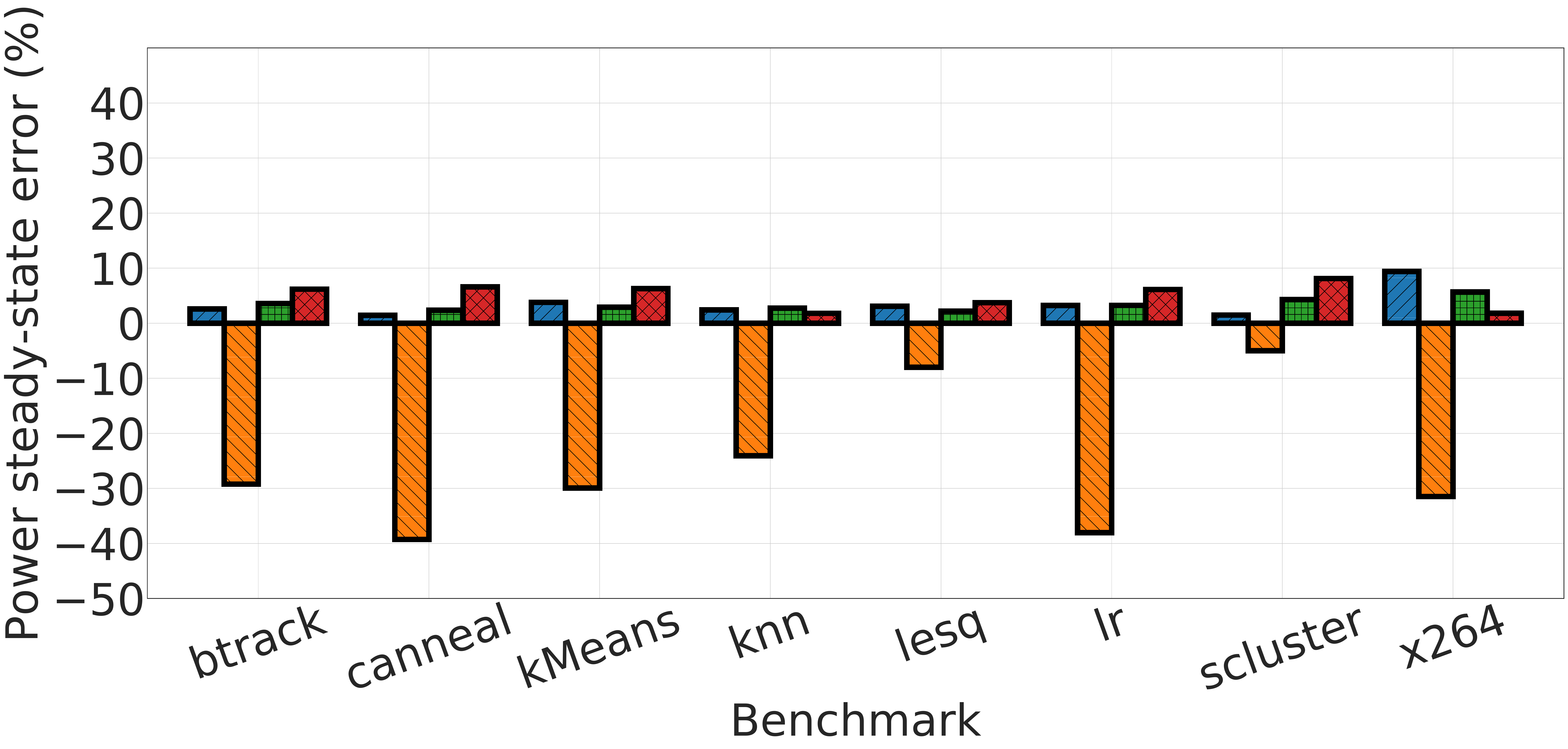}
                \label{fig:results:ss:ph3:pow}
        }%
        \caption{Steady-state error for all benchmarks, grouped by phase. A negative value indicates the amount of power/QoS \textbf{exceeding} the reference value (bad), a positive value indicates the amount of power saved (good) or QoS degradation (bad) (from \cite{Rahmani2018}).}
        \label{fig:results:saso:ss}
\end{figure*}

\paragraph{\texttt{x264} Benchmark}
To show the energy efficiency of \supervisory, we study the Safe Phase.
The Safe Phase consists of the first 5 seconds of execution during which only the QoS application executes on the Big cluster.
In this phase, all controllers are able to achieve the FPS reference value within the power envelope.
Figures \ref{fig:results:ss:ph1:qos} and \ref{fig:results:ss:ph1:pow} show the average steady-state error (\%) of QoS and power respectively for each resource manager in Phase 1.
Steady-state error is used to define \emph{accuracy} in feedback control systems~\cite{Hellerstein:2004:FCC}.
Steady-state error values are calculated as $\mathit{reference} - \mathit{measured~output}$.
Negative values indicate that the power/QoS \textbf{exceeds} the reference value, positive values indicate power savings or failure to meet QoS.
We make two key observations.
First, both \multiPerf{} and \supervisory{} reduce power consumption by 25\% (Fig.~\ref{fig:results:ss:ph1:pow}) while maintaining FPS within 10\% (Fig.~\ref{fig:results:ss:ph1:qos}) of the reference value.
The \multiPerf{} controller operates efficiently because the reference FPS value is achievable within the TDP threshold.
The \supervisory{} controller similarly operates efficiently: it is able to recognize that the FPS is achievable within TDP and, as a result, lower the reference power.
Second, the \fs{} and \multiPow{} controllers unnecessarily exceed the reference FPS value and, as a result, consume excessive power.
This is because these controllers prioritize meeting the power reference value, consuming the entire available power budget to maximize performance.

To show \supervisory's ability to adapt to a sudden change in operating constraints, we study the Emergency Phase.
The Emergency Phase of execution emulates a thermal emergency, during which, the TDP is lowered to ensure that the system operates in a safe state.
This occurs during the second 5-second period of execution in Figure~\ref{fig:results:time}.
We observe that all controllers are able to react to the change in power reference value and maintain QoS.
However, compared to the other controllers, \fs{} has a sluggish reaction (Figure~\ref{fig:results:fs:power}) to the change in power reference, despite the fact that it is designed to prioritize tracking the power output.
\emph{Settling time} is a property used to quantify responsiveness of feedback control systems~\cite{Hellerstein:2004:FCC}.
\emph{Settling time} is the time it takes to reach sufficiently close to the steady-state value after the reference values are set.
The average settling time for the power output of \fs{} is 2.07 seconds, while \supervisory{} has an average settling time of 1.28 seconds.
The larger size of the state-space (\textit{x(t)} matrix in Equation \ref{eq.lqg} and \ref{eq.lqg2}) and the higher number of control inputs in the 4$\times$2 \fs{} compared to those of 2$\times$2 controllers in \supervisory{} is the reason for the slow settling time of \fs. 
This is also the reason why SISO controllers are generally faster that MIMOs \cite{Hellerstein:2004:FCC}.

To show \supervisory's ability to adapt to workload disturbance and changing system goals, we study the Workload Disturbance Phase.
The Workload Disturbance Phase occurs in seconds 10-15 of execution in Figure~\ref{fig:results:time}.
In this phase, 1) the QoS reference value and the power envelope return to the same values as in Phase 1, and 2) we introduce disturbance in the form of background tasks.
As a result of the workload disturbance, the QoS reference is \emph{not} achievable within the TDP.
We make two observations regarding the steady-state error in Figures~\ref{fig:results:ss:ph3:qos} and \ref{fig:results:ss:ph3:pow}.
First, \supervisory{} behaves similarly to \multiPow, even though in Phase 1 it behaved similarly to \multiPerf.
The \supervisory{} supervisor is able to recognize the change in execution scenario and constraints, and adapt its priorities appropriately.
In this case, \supervisory{} achieves much higher FPS than all controllers except \multiPerf{} (Fig.~\ref{fig:results:ss:ph3:qos}), while obeying the TDP limit (Fig.~\ref{fig:results:ss:ph3:pow}).
Second, both \fs{} and \multiPow{} operate at the TDP limit, but achieve a significantly lower FPS than the reference value.
\multiPerf{} comes within \textasciitilde5\% of the reference FPS (Fig.~\ref{fig:results:ss:ph3:qos}) while exceeding the TDP by more than 30\% (Fig.~\ref{fig:results:ss:ph3:pow}), which is undesirable.

\paragraph{Other Benchmarks}
We perform the same experiments for PARSEC benchmarks \texttt{bodytrack},
\texttt{canneal}, \texttt{streamcluster}, as well as machine-learning benchmarks \texttt{k-means}, \texttt{KNN}, \texttt{least} \texttt{squares}, and \texttt{linear regression}.
For these workloads, we use the generic \emph{heartbeat rate} (HB) directly as the QoS metric, as FPS is not an appropriate metric.
Figures \ref{fig:results:ss:ph1:qos}, \ref{fig:results:ss:ph2:qos}, and \ref{fig:results:ss:ph3:qos} show the average steady-state error (\%) of QoS for Phases 1, 2, and 3 respectively.
Figures \ref{fig:results:ss:ph1:pow}, \ref{fig:results:ss:ph2:pow}, and \ref{fig:results:ss:ph3:pow} show the average steady-state error (\%) of power for Phases 1, 2, and 3 respectively.
We summarize the observations for the additional experiments with respect to \texttt{x264} for the three phases.
In the Safe Phase, the behavior of \texttt{bodytrack}, \texttt{streamcluster}, \texttt{k-means}, \texttt{KNN}, \texttt{least squares}, and \texttt{linear regression} is similar to that of \texttt{x264} (Figures~\ref{fig:results:ss:ph1:qos} and \ref{fig:results:ss:ph1:pow}).
\texttt{canneal} follows the same pattern with respect to power as all other benchmarks (Fig.~\ref{fig:results:ss:ph1:pow}).
\texttt{canneal}'s QoS steady-state error is the only difference in behavior we observe in Phase 1.
None of the managers are able to meet the QoS reference value for \texttt{canneal} in Phase 1 (Fig.~\ref{fig:results:ss:ph1:qos}).
This is due to the fact that the phase of \texttt{canneal} captured in the experiment primarily consists of serialized input processing, so the number of idle cores has reduced affect on QoS.
In the Emergency Phase, our observations from \texttt{x264} hold for nearly all benchmarks regarding response to change in power reference value, achieving less than 10\% power steady-state error (Fig.~\ref{fig:results:ss:ph2:pow}). 
The only exceptions are \texttt{canneal} and \texttt{k-means}: the \multiPerf{} manager is unable to react to change in TDP for \texttt{canneal} and \texttt{k-means}.
The \multiPerf{} manager lacks a supervisory coordinator and prioritizes performance, and was unable to find a configuration for \texttt{canneal} and \texttt{k-means} that satisfied the QoS reference value within TDP.
In the Workload Disturbance Phase, \supervisory, \fs, and \multiPow{} all achieve near-reference power (Fig.~\ref{fig:results:ss:ph3:pow}).
As expected, \multiPerf{} violates the TDP in all cases, but always achieves the highest QoS (Fig.~\ref{fig:results:ss:ph3:qos}).

We conclude that \supervisory{} is effective at (1) efficiently meeting multiple system objectives when it is possible to do so, (2) appropriately balancing multiple conflicting objectives, and (3) quickly responding to sudden and unpredictable changes in constraints due to workload or system state.

\subsubsection{Overhead Evaluation}
To show the overhead of the low-level MIMO controllers, we study their execution time.
We measure the MIMO controller execution time to be $2.5ms$, on average, over 30 seconds.
The MIMO controller is invoked every 50\textit{ms} resulting in a 5\% overhead, which is experienced by all evaluated controllers.
We measure the runtime of the supervisor to be $30\mu s$, which is negligible even with respect to the MIMO controller execution time.
The supervisor is invoked less frequently than the MIMO controllers ($2\times$ the period in our case), executes in parallel to the workload and MIMO controllers, and simply evaluates the system state in order to determine if the MIMO controller gains need changing. 
State changes that result in interventions on the low-level controllers occur only due to system-wide changes in the state (e.g., thermal emergency) or goals (e.g., change in performance reference value or execution mode), which are infrequent.
When the supervisor needs to change the MIMO gains, it simply points the coefficient matrices to a different set of stored values.
In our case study, we have two sets of gains (QoS and power oriented) that are generated when the controllers are designed and stored during system initialization.
Changing the coefficient arrays at runtime takes effect immediately, and has no additional overhead.

To show the overhead of \supervisory's supervisory controller, we compare the total execution time of identical workloads with and without \supervisory.
With respect to the preemption overhead due to globally managing resources, Linux's HMP scheduler typically maps SCT threads to a core on the low-power Little cluster.
Therefore, the SCT threads are executed without preempting the QoS application, which always executes on the Big cluster. 
We verify the overall impact of the control system overhead by running the benchmarks on two different systems: i) a vanilla Linux setup\footnote{
Ubuntu 16.04.2 LTS and Linux kernel 3.10.105 (https://dn.odroid.com/5422/ODROID-XU3/Ubuntu/).}
and ii) vanilla Linux with \supervisory{} running in the background. 
For (ii), \supervisory{} controllers perform all the required computations but do \emph{not} change the system knobs (thus only the \supervisory{} overhead affects the system). 
When comparing the QoS of the applications across multiple runs, we verify a negligible average difference of 0.1\% between the two systems.

We conclude that the benefits of \supervisory{} come at a negligible performance overhead.

\subsection{Summary}
Modern mobile systems require intelligent management to balance user demands and system constraints.
At any given time, the relative priority of demands and constraints may change based on uncontrollable context, such as dynamic workload or operating condition. 
A resource manager must be able to autonomously detect such context changes and adapt appropriately. 
This property is known as self-adaptivity. 
We demonstrate one way to design a self-adaptive resource manager: using supervisory control theory. 
Supervisory control theory lends itself well to this challenge due to its high level of abstraction and lightweight implementation. 
The proposed supervisor successfully adapts to changes when managing quality of service under a power budget for chip multiprocessors.
The hierarchy using supervisory control theory represents early exploration of self-adaptivity in the resource management domain, and a slight degree of self-awareness. 
This approach can be enhanced in one way through the definition and generation of goals. 
Initial work based on goaldriven autonomy has been done toward this end \cite{shamsa}.

\section{Conclusion}

We use two forms of computational self-awareness to implement resource managers with simple but effective self-aware components.
Systems can be self-aware to varying degrees, and the degree to which self-\emph{X} properties are utilized is case-specific.
We demonstrate the use of self-optimization in implementing a DVFS governor for managing power in a processor core.
We demonstrate the use of self-adaptivity in implementing a multi-goal resource manager for managing QoS within a power budget in an HMP.
Moving forward, as systems scale and configuration spaces grow, computational self-awareness provides a useful abstract tool for tackling various challenges in resource management.

\begin{acknowledgement}
This work was partially supported by NSF grant CCF-1704859.
\end{acknowledgement}


\bibliographystyle{spbasic}
\bibliography{refs}

\end{document}

%% file: fig/feedback_controller_reflection.tex
\begin{tikzpicture}[font={\scriptsize}, >=latex, every text node part/.style={align=center}]
	\def\mwdth{2.5cm}
	\def\mhgth{1.0cm}
	\def\mdis{0.75cm}

	\node [draw, minimum width=\mwdth, minimum height=\mhgth, thick] (Controller) at (0,0) {Controller\\\textit{\textbf{Decide}}};
	\node [draw, color = blue, minimum width=\mwdth, minimum height=\mhgth*0.5, above = 0 of Controller] (Prediction) {\textit{\textbf{Predict}}};
	\node [draw, minimum width=\mwdth, minimum height=\mhgth, right = 3*\mdis of Controller] (System) {System};
	\node [circle, draw, minimum width=1*\mhgth, minimum height=1*\mhgth, left = \mdis of Controller] (Diff) {};
 	\node [] () at ($(Diff) - (0,0)$) {---};
	\draw[->] (Diff) -- (Controller);
	\draw[->] (Controller) -- (System) node [midway, below] (Act_Freq) {\textit{\textbf{Act}}} ;

	\node [left= 0.1*\mdis of Diff, opacity=0.0] (tmp) {Target};
	\node [rotate around={90:(tmp.center)}] (Ref) at (tmp) {Target};
	\draw[->] (Ref) -- (Diff);
	\coordinate[] (FBhelp) at ($(System) + (5*\mdis, -2*\mdis)$);
	\draw[->] (System) -| node [near start, above] (Sen_Heart) {\textit{\textbf{Observe}}} node [near start, below] (Sen_IPS) {} (FBhelp) -| (Diff);
	\node[cloud, cloud puffs=10.8,cloud puff arc=110, aspect=2, draw, color = blue, above = \mdis of Act_Freq 
    ] (Reflection)  {Reflection};
	\coordinate[] (Refhelp) at ($(System) + (5*\mdis, 0)$);
    \draw[->, thick, color=blue] (Refhelp) |- (Reflection);
    \draw[->, thick, color=blue] (Reflection) -| (Prediction);
    \draw[->, thick, color=blue] (Act_Freq) -- (Reflection);
\end{tikzpicture}

%% file: fig/arm_big_little_cluster_f-p.tex
\begin{tikzpicture}[scale = 0.90, font=\footnotesize]
        \begin{axis}[ 
            xlabel={Frequency (MHz)}, 
            ylabel={Power (W)},
            legend pos=north west,
            width=\linewidth,
            height=4cm,
            legend style={font=\footnotesize}
        ] 
        \addplot+[magenta,mark options={fill=magenta}] coordinates { 
            (2000,9.4) (1900,7.6) (1800,6.6) (1700,5.7) (1600,5.0)
            (1500,4.3) (1400,3.8) (1300,3.4) (1200,3.0) (1100,2.6)
            (1000,2.2) (900,1.9) (800,1.6) (700,1.3) (600,1.1)
            (500,1.0) (400,0.8) (300,0.6) (200,0.4)
        }; 
        \addplot+[brown,mark options={fill=brown}] coordinates{ 
            (1400,0.42) (1300,0.36) (1200,0.30) (1100,0.26)
            (1000,0.22) (900,0.20) (800,0.14) (700,0.12) (600,0.10)
            (500,0.08) (400,0.07) (300,0.06) (200,0.04)
        }; 
        \legend{ARM A15 MP,ARM A7 MP} 
        \end{axis} 
    \end{tikzpicture}

%% file: fig/gsc_motivation_time_full.tex
\begin{tikzpicture}[font=\footnotesize, transform shape]
		
	\begin{axis}[width=\linewidth, height=0.5\linewidth, xlabel=Time (seconds), ylabel=Power (W), axis lines=left, no markers, grid=major, legend pos=north east, xmin=0, xmax=30, ymin=0, ymax=2.5, ytick distance = 0.5, legend style={font=\footnotesize, legend columns=-1}] 

			\fill [yellow] (axis cs:0,1.3) rectangle (axis cs:10,1.7);

	    \addplot [magenta] table [x=time, y=power_w_filtered, col sep = comma] {data/motivation_time_full.csv };
	    
	    \addplot [black,dashed, thick] table [x=time, y=power_ref, col sep = comma ] {data/motivation_time_full.csv };
	    
  		\addlegendentry[]{Power Measured}
  		\addlegendentry[]{Power Reference}

	\end{axis}

\end{tikzpicture}

%% file: fig/gsc_motivation_time_l2l3.tex
\begin{tikzpicture}[font=\footnotesize, transform shape]
		
	\begin{axis}[width=\linewidth, height=0.5\linewidth, xlabel=Time (seconds), ylabel=Power (W), axis lines=left, no markers, grid=major, legend pos=north east, xmin=0, xmax=30, ymin=0, ymax=2.5, ytick distance = 0.5, legend style={font=\footnotesize,legend columns=-1}] 

			\fill [yellow] (axis cs:0,1.3) rectangle (axis cs:10,1.7);

	    \addplot [magenta] table [x=time, y=power_w_filtered, col sep = comma] {data/motivation_time_l2l3.csv };
	    
	    \addplot [black,dashed, thick] table [x=time, y=power_ref, col sep = comma ] {data/motivation_time_l2l3.csv };
	    
  		\addlegendentry[]{Power Measured}
  		\addlegendentry[]{Power Reference}

	\end{axis}

\end{tikzpicture}

%% file: tab/a15_v-f_pairs.tex
  \begin{tabular}{c|c|c} 
  \hline
  Region & Frequency & Voltage \\ & Range (MHz) & (V) \\ 
  \hline
    1 & 1600 -- 2000 & 1.25  \\
    2 & 1300 -- 1500 & 1.10  \\
    3 & 900 -- 1200 & 1.00  \\
    4 & 200 -- 800 & 0.90
  \end{tabular}

%% file: fig/gsc_block_diagram.tex
\begin{tikzpicture}[font={\scriptsize}, >=latex, every text node part/.style={align=center}]
\def\mwdth{1.25cm}
\def\mhgth{0.5cm}
\def\mdis{0.5cm}

\newcommand\ppbb{path picture bounding box}
\tikzset{
switch/.style = {minimum size=3em,
	path picture={  \draw
		([xshift=-3mm]  \ppbb.center)  -- ++ (45:6mm);
		\draw[shorten >=3mm,-]
		(\ppbb.west) edge (\ppbb.center);
		\draw[<->]
		([yshift=-2mm] \ppbb.center) arc[start angle=-15, end angle=75, radius=6mm];   
	},
	label={[yshift=-2mm] above:#1},
	node contents={}},
}

\node [draw, minimum width=\mwdth, minimum height=\mhgth*2, thick] (Controller) at (0,0) {Controller} node [below = 0mm of Controller] {\textit{\textbf{Reason}}};

\node [draw, minimum width=\mwdth, minimum height=\mhgth, right = 3*\mdis of Controller] (System) {System};

\node [circle, draw, minimum width=1*\mhgth, minimum height=1*\mhgth, left = \mdis of Controller] (Diff) {};
\node [] () at ($(Diff) - (0,0)$) {+};
\node [] () at ($(Diff) - (0,0.25*\mhgth)$) {$-$};

\draw[->,thick] (Diff) -- (Controller);
\draw[->,thick] (Controller) -- (System) node [midway, below] (Act_Freq) {\textit{\textbf{Act}}} ;

\node [left= \mdis of Diff, opacity=0.0] (tmp) {Goal};
\node (Ref) at (tmp) {\textit{Goal}};
\draw[->,thick] (Ref) -- (Diff);

\coordinate[] (FBhelp) at ($(System) + (5*\mdis, -2*\mdis)$);
\draw[->,thick] (System) -| node [near start, above] (Sen_Heart) {\textit{\textbf{Sense}}} node [near start, below] (Sen_IPS) {} (FBhelp) -| (Diff);

	\node[draw, above = 2*\mdis of Act_Freq, minimum width = \mwdth] (GAINSN)  {Gains $N$};
	\node[draw, circle, above = 1mm of GAINSN, minimum width = 1mm, inner sep = 0mm] (DOT1)  {};
	\node[draw, circle, above = 1mm of DOT1, minimum width = 1mm, inner sep = 0mm] (DOT2)  {};
	\node[draw, circle, above = 1mm of DOT2, minimum width = 1mm, inner sep = 0mm] (DOT3)  {};
	\node[draw, above = 1mm of DOT3, minimum width = \mwdth] (GAINS1)  {Gains 1};
	\coordinate[above = \mdis of System] (GainsHelp1);
	\coordinate[right = 2*\mdis of DOT2] (GainsHelp2);
	\draw[-, thick, color=blue] (Act_Freq) |- (GainsHelp1) node [below] (X) {Scheduling Variable};
	\draw[->, thick, color=blue] (GainsHelp1) |- (GAINS1);
	\draw[->, thick, color=blue] (GainsHelp1) |- (GAINSN);
	\coordinate[left = \mdis of GAINS1] (LGAIN1);
	\draw [-]  (GAINS1) -- (LGAIN1);
	\node at (LGAIN1.west) [circle,fill,inner sep=0mm, minimum width = 1mm]{};
	\coordinate[left = \mdis of GAINSN] (LGAINN);
	\draw [-]  (GAINSN) -- (LGAINN);
	\node at (LGAINN.west) [circle,fill,inner sep=0mm, minimum width = 1mm]{};
	
	\node (Switch)      [switch, left= \mdis+3mm of DOT2] {};
	\coordinate[left = 0mm of Switch] (SwitchHelp1);
	\coordinate[above = \mdis of Controller] (SwitchHelp2);
	\draw [-, thick, color=blue]  (SwitchHelp1) |- (SwitchHelp2) node [below left] (X) {Control Parameters};
	\draw [->, thick, color=blue]  (SwitchHelp2) -- (Controller);

	\node[above , color=blue] at (GAINS1.north west) (GSClabel) {Gain Scheduler};
	    \node[draw, dashed, thick, color=blue, inner sep=1mm, fit=(GAINS1) (GAINSN) (Switch) (GSClabel)] (GSC) {};

\end{tikzpicture}

%% file: tab/dvfs_siso_accuracy.tex
  \begin{tabular}{l||r|r|r|r|}
  \hline
    & Ctrl 1 & Ctrl 2.1 & Ctrl 2.2 & Ctrl 2.3 \\ 
  \hline
    Freq. Range & 200 -- 1800 & 1300 -- 1800 & 900 -- 1200 & 200 -- 800 \\
  \hline
    Stable & \checkmark  & \checkmark & \checkmark  & \checkmark \\
  \hline
    \makecell[cl]{Accuracy \\ (MSE)} & 0.1748 & 0.03089 & 0.0005382 & 0.0003701  \\
  \hline
  \end{tabular}

%% file: fig/results/gsc-tot_os.tex
\pgfplotstableread[col sep=comma,]{data/c1-date18.csv}\datatableC
\pgfplotstableread[col sep=comma,]{data/c1-date18.csv}\datatableCB
\pgfplotstableread[col sep=comma,]{data/gsc-date18.csv}\datatableGSC

\begin{tikzpicture}[transform shape, font=\huge]

	\begin{axis}[ybar, ymin=0,  xmin=0, ylabel=Power \emph{over} ref. (W), ylabel near ticks, xlabel=Benchmark, xticklabels={btrack,scluster,x264}, xtick=data, enlarge x limits=0.25, width=\textwidth, xticklabel style={rotate=45}, xlabel near ticks,tick pos=left, legend style={font=\LARGE}, bar width = 1cm]
		\addplot[magenta, fill=magenta] table [x expr=\coordindex, y=tot_os, col sep = comma] {\datatableC};
		\addlegendentry{SISO}

		\addplot[brown,fill=brown] table [x expr=\coordindex, y=tot_os, col sep = comma] {\datatableGSC};
		\addlegendentry{GSC}
	\end{axis}

\end{tikzpicture}

%% file: fig/results/gsc-tot_us.tex
\pgfplotstableread[col sep=comma,]{data/c1-date18.csv}\datatableC
\pgfplotstableread[col sep=comma,]{data/c1-date18.csv}\datatableCB
\pgfplotstableread[col sep=comma,]{data/gsc-date18.csv}\datatableGSC

\begin{tikzpicture}[transform shape, font=\huge]

	\begin{axis}[ybar, ymin=0,  xmin=0, ylabel=Power \emph{under} ref. (W), ylabel near ticks, xlabel=Benchmark, xticklabels={btrack,scluster,x264}, xtick=data, enlarge x limits=0.25, width=\textwidth, xticklabel style={rotate=45}, xlabel near ticks,tick pos=left, legend style={font=\LARGE}, bar width = 1cm]
		\addplot[magenta,fill=magenta] table [x expr=\coordindex, y=tot_us, col sep = comma] {\datatableC};
		\addlegendentry{SISO}

		\addplot[brown,fill=brown] table [x expr=\coordindex, y=tot_us, col sep = comma] {\datatableGSC};
		\addlegendentry{GSC}
	\end{axis}

\end{tikzpicture}

%% file: fig/results/gsc-act.tex
\pgfplotstableread[col sep=comma,]{data/c1-date18.csv}\datatableC
\pgfplotstableread[col sep=comma,]{data/c1b-date18.csv}\datatableCB
\pgfplotstableread[col sep=comma,]{data/gsc-date18.csv}\datatableGSC

\begin{tikzpicture}[transform shape, font=\huge]

	\begin{axis}[ybar, ymin=0,  xmin=0, ylabel=Total \# Actuations, ylabel near ticks, xlabel=Benchmark, xticklabels={btrack,scluster,x264}, xtick=data, enlarge x limits=0.25, width=\textwidth, xticklabel style={rotate=45}, xlabel near ticks,tick pos=left, 
	legend style={font=\LARGE}, bar width = 1cm
	]
		\addplot[magenta,fill=magenta] table [x expr=\coordindex, y=tot_act, col sep = comma] {\datatableC};
		\addlegendentry{SISO}

	\addplot[brown,fill=brown] table [x expr=\coordindex, y=tot_act, col sep = comma] {\datatableGSC};
	\addlegendentry{GSC}

		\addplot[orange,fill=orange] table [x expr=\coordindex, y=tot_act, col sep = comma] {\datatableCB};
		\addlegendentry{SISO A}
	\end{axis}

\end{tikzpicture}

%% file: fig/results/gsc-st.tex
\pgfplotstableread[col sep=comma,]{data/c1-date18.csv}\datatableC
\pgfplotstableread[col sep=comma,]{data/c1-date18.csv}\datatableCB
\pgfplotstableread[col sep=comma,]{data/gsc-date18.csv}\datatableGSC

\begin{tikzpicture}[ transform shape, font=\huge]

	\begin{axis}[ybar, ymin=0,  xmin=0, ylabel=Settling Time (ms), ylabel near ticks, xlabel=Benchmark, xticklabels={btrack,scluster,x264}, xtick=data, enlarge x limits=0.25, width=\textwidth, xticklabel style={rotate=45}, xlabel near ticks,tick pos=left, legend style={font=\LARGE}, legend pos=north west, bar width = 1cm]
		\addplot[magenta,fill=magenta] table [x expr=\coordindex, y=st_ms, col sep = comma] {\datatableC};
		\addlegendentry{SISO}

		\addplot[brown,fill=brown] table [x expr=\coordindex, y=st_ms, col sep = comma] {\datatableGSC};
		\addlegendentry{GSC}
	\end{axis}

\end{tikzpicture}

%% file: fig/MIMO_2x2.tex
\begin{tikzpicture}[font={\normalsize}, >=latex, every text node part/.style={align=center}]
	\def\mwdth{2.5cm}
	\def\mhgth{1.0cm}
	\def\mdis{1.0cm}
	\node [draw, minimum width= \mwdth, minimum height= 2*\mhgth, thick, outer sep = 0] (Controller) at (0,0) {Controller};
	
	\node [draw, minimum width= \mwdth, minimum height= 2*\mhgth, right = 3*\mdis of Controller] (System) {System};
	
	\node [circle, draw, minimum width= \mhgth, minimum height= \mhgth, above left = 0.25*\mhgth and 2*\mdis of Controller.west] (Diff) {};
	\node [] () at ($(Diff) - (0.75*\mhgth*0.5,0)$) {+};
	\node [] () at ($(Diff) - (0,0.75*\mhgth*0.5)$) {--};
	\node [circle, draw, minimum width= \mhgth, minimum height= \mhgth, below right = 0.5*\mhgth and 0.5*\mdis of Diff] (Diff2) {};
	\node [] () at ($(Diff2) - (0.75*\mhgth*0.5,0)$) {+};
	\node [] () at ($(Diff2) - (0,0.75*\mhgth*0.5)$) {--};
	
	\node [above = 0.5*\mhgth of Controller.west] (Cin1) {};
	\draw[->] (Diff) -- (Cin1);
	\node [below = 0.5*\mhgth of Controller.west] (Cin2) {};
	\draw[->] (Diff2) -- (Cin2);
	
	\node [above = 0.5*\mhgth of Controller.east] (Cout1) {};
	\node [below = 0.5*\mhgth of Controller.east] (Cout2) {};
	\node [above = 0.5*\mhgth of System.west] (Sin1) {};
	\node [below = 0.5*\mhgth of System.west] (Sin2) {};
	
	\draw[->] (Cout1) -- (Sin1) node [midway, above] (Act_Freq) {Frequency} node [midway, below] (Act_Voltage) {\footnotesize(Voltage)};
	\draw[->] (Cout2) -- (Sin2) node [midway, above] (Act_Cores) {Idle Cores} node [midway, below] (Act_Cores) {};
	
	\node[draw, dashed, inner sep=0.1mm, fit=(Act_Freq) (Act_Cores), label=below:{\textcolor{gray}{\footnotesize Actuators}}, label=above:{\textcolor{gray}{\footnotesize Control Input}}] (Actuators) {};
	
	\node [left= 0.1*\mdis of Diff, opacity=0.0] (tmp) {Reference\\FPS};
	\node [rotate around={90:(tmp.center)}] (Ref) at (tmp) {Reference\\FPS};
	\draw[->] (Ref) -- (Diff);
	
	\node [left= 0.7*\mdis of Diff2, opacity=0.0] (tmp) {Reference\\Power};
	\node [rotate around={90:(tmp.center)}] (Ref2) at (tmp) {Reference\\Power};
	\draw[->] (Ref2) -- (Diff2);
	
	\node [above = 0.5*\mhgth of System.east] (Sout1) {};
	\node [below = 0.5*\mhgth of System.east] (Sout2) {};
	\coordinate[] (FBhelp) at ($(System) + (5*\mdis, -2*\mdis)$);
	\coordinate[] (FBhelp2) at ($(System) + (4.5*\mdis, -1.5*\mdis)$);
	\draw[->] (Sout1) -| node [near start, above] (Sen_Frames) {Frames-Per-Second} node [near start, below] (Sen_FPS) {\footnotesize (FPS)} (FBhelp) -| (Diff);
	\draw[->] (Sout2) -| node [near start, above] (Sen_Pow) {Power} node [near start, below] (Sen_Pow) {} (FBhelp2) -| (Diff2);
	
	\node[draw, dashed, inner sep=0.1mm, fit= (Sen_Frames) (Sen_FPS) (Sen_Pow), label=below:{\textcolor{gray}{\footnotesize Sensors}}, label=above:{\textcolor{gray}{\footnotesize Measured Output}}] (Sensors) {};
	
\end{tikzpicture}

%% file: fig/sct_overview.tex
\begin{tikzpicture}[scale = 1.5, node distance = 1.5cm, font={\scriptsize}, >=latex, every text node part/.style={align=center}]
	\def\mwdth{2cm}
	\def\mdis{0.5cm}
	\node [draw, fill=white, minimum width=\mwdth, minimum height=0.5cm,] (LLC1) {Low-level\\Controller};
	\node [draw, fill=white, minimum width=\mwdth, minimum height=0.5cm, right = 2*\mdis of LLC1] (SUBSYS1) {Subsystem\\};
	\node [draw, minimum width=\mwdth, minimum height=0.5cm, above = 2*\mdis of LLC1] (SC) {Supervisory\\Controller};
	\node [draw, minimum width=\mwdth, minimum height=0.5cm, right = 2*\mdis of SC] (MODEL) {System\\Model};
	
	\draw [thin, fill=white, -Implies, double distance=1.5pt] (SC) -- (LLC1);
	\node [below left = \mdis/2-2mm and -0.6cm of SC] (ctrl_params) {\tiny \begin{parbox}{0.5cm}{ \texttt{Control\\Parameters}}\end{parbox}};

	\draw [thin, fill=white, Implies-, double distance=1.5pt] (MODEL) -- (SUBSYS1);
	\node [below left = \mdis/2-2mm and -0.8cm of MODEL] (ctrl_params) {\tiny \begin{parbox}{0.5cm}{ \texttt{Model\\Updates}}\end{parbox}};

	\coordinate[below left = -2mm and 0mm of MODEL] (help0);
	\coordinate[below right = -2mm and 0mm of SC] (help1);
	\draw [thin, fill=white, -Implies, double distance=1.5pt] (help0) -- (help1);
	\node [below left = -0.1cm and \mdis of MODEL] (ctrl_params) {\tiny \begin{parbox}{0.5cm}{ \texttt{Information}}\end{parbox}};

	\coordinate[above left = -2mm and 0mm of MODEL] (help2);
	\coordinate[above right = -2mm and 0mm of SC] (help3);
	\draw [thin, fill=white, -Implies, dashed, double distance=1.5pt] (help3) -- (help2);
	\node [above right = -0.2cm and \mdis/2-1mm of SC] (ctrl_params) {\tiny \begin{parbox}{0.5cm}{ \texttt{Virtual\\Control}}\end{parbox}};

	\coordinate[below left = -2mm and 0mm of SUBSYS1] (help4);
	\coordinate[below right = -2mm and 0mm of LLC1] (help5);
	\draw [thin, fill=white, -Implies, double distance=1.5pt] (help4) -- (help5);
	\node [below left = -0.1cm and \mdis/2 of SUBSYS1] (ctrl_params) {\tiny \begin{parbox}{0.5cm}{ \texttt{Feedback}}\end{parbox}};

	\coordinate[above left = -2mm and 0mm of SUBSYS1] (help6);
	\coordinate[above right = -2mm and 0mm of LLC1] (help7);
	\draw [thin, fill=white, -Implies, double distance=1.5pt] (help7) -- (help6);
	\node [above right = 0cm and \mdis/2 of LLC1] (ctrl_params) {\tiny \begin{parbox}{0.5cm}{ \texttt{Control}}\end{parbox}};

	
	\begin{pgfonlayer}{bg1}    
        \node [draw, fill=white, minimum width=\mwdth, minimum height=0.5cm, above right = -0.6cm and -1.9cm of SUBSYS1] (SUBSYS2) {Subsystem\\};
        \node [draw, fill=white, minimum width=\mwdth, minimum height=0.5cm, above right = -0.6cm and -1.9cm of LLC1] (LLC2) {Low-level\\Controller};
    	
    	\coordinate[below left = 0mm and -\mwdth/2-0.01cm of SC] (help);
    	\draw [thin, fill=black, -Implies, double distance=1.5pt] (help) -- (LLC2);

        \coordinate[below left = 0mm and -\mwdth/2-0.01cm of MODEL] (help);
    	\draw [thin, fill=black, Implies-, double distance=1.5pt] (help) -- (SUBSYS2);
	
    	\coordinate[below left = -2mm and 0mm of SUBSYS2] (help4);
    	\coordinate[below right = -2mm and 0mm of LLC2] (help5);
    	\draw [thin, fill=white, -Implies, double distance=1.5pt] (help4) -- (help5);
    	
    	\coordinate[above left = -2mm and 0mm of SUBSYS2] (help6);
    	\coordinate[above right = -2mm and 0mm of LLC2] (help7);
    	\draw [thin, fill=white, -Implies, double distance=1.5pt] (help7) -- (help6);
	\end{pgfonlayer}	
	
	\begin{pgfonlayer}{bg2}    
        \node [draw, minimum width=\mwdth, minimum height=0.5cm, above right = -0.6cm and -1.9cm of SUBSYS2] (SUBSYS3) {Subsystem\\};
	    \node [draw, fill=white, minimum width=\mwdth, minimum height=0.5cm, above right = -0.6cm and -1.9cm of LLC2] (LLC3) {Low-level\\Controller};
	    
    	\coordinate[below left = 0mm and -\mwdth/2-0.02cm of SC] (help);
    	\draw [thin, -Implies, double distance=1.5pt] (help) -- (LLC3);

        \coordinate[below left = 0mm and -\mwdth/2-0.02cm of MODEL] (help);
    	\draw [thin, Implies-, double distance=1.5pt] (help) -- (SUBSYS3);
	
    	\coordinate[below left = -2mm and 0mm of SUBSYS3] (help4);
    	\coordinate[below right = -2mm and 0mm of LLC3] (help5);
    	\draw [thin, fill=white, -Implies, double distance=1.5pt] (help4) -- (help5);
    	
    	\coordinate[above left = -2mm and 0mm of SUBSYS3] (help6);
    	\coordinate[above right = -2mm and 0mm of LLC3] (help7);
    	\draw [thin, fill=white, -Implies, double distance=1.5pt] (help7) -- (help6);
    \end{pgfonlayer}
	
	\node[fill=gray!40, fill opacity=0.2, draw, dashed, thick, inner sep=2mm, fit=(SUBSYS1) (SUBSYS2) (SUBSYS3), label=below:\textbf{\color{gray}System}] {};
	\node[fill=gray!40, fill opacity=0.2, draw, dashed, thick, inner sep=2mm, fit=(LLC1) (LLC2) (LLC3), label=below:\textbf{\color{gray}Leaf Controllers}] {};
\end{tikzpicture}

%% file: fig/sct_gsc.tex
\begin{tikzpicture}[node distance = 1.5cm, font={\scriptsize}, >=latex, every text node part/.style={align=center}]
	\def\mwdth{2cm}
	\def\mdis{0.5cm}

	\begin{scope}[shift={(0,0)}]{
		\def\mwdth{1.25cm}
		\def\mhgth{0.5cm}
		\def\mdis{0.5cm}
			
		\node [draw, minimum width=\mwdth, minimum height=\mhgth*2, thick] (Controller) at (0,0) {Controller} node [below = 0mm of Controller] (Reason) {\textit{\textbf{Reason}}};
		
		\node [draw, minimum width=\mwdth, minimum height=\mhgth, right = 3*\mdis of Controller] (System) {Subsystem};
		\node [circle, draw, minimum width=1*\mhgth, minimum height=1*\mhgth, left = \mdis of Controller] (Diff) {};
		\node [] () at ($(Diff) - (0,0)$) {+};
		\node [] () at ($(Diff) - (0,0.25*\mhgth)$) {$-$};
		\draw[->,thick] (Diff) -- (Controller);
		\draw[->,thick] (Controller) -- (System) node [midway, below] (Act_Freq) {\textit{\textbf{Act}}} ;

		\node [left= \mdis of Diff, opacity=0.0] (tmp) {Goal};
		\node (Ref) at (tmp) {\textit{Goal}};
		\draw[->,thick] (Ref) -- (Diff);

		\coordinate[] (FBhelp) at ($(System) + (5*\mdis, -2*\mdis)$);
		\draw[->,thick] (System) -| node [near start, above] (Sen_Heart) {\textit{\textbf{Sense}}} node [near start, below] (Sen_IPS) {} (FBhelp) -| (Diff);	
	}
	\end{scope}

	\node[state,fill=gray,thick, minimum size = 3mm]         (A) [above = 4*\mdis of Controller]             {};
	\node[state,fill=gray,thick, minimum size = 3mm]         (B) [below right = 3mm of A] {};
	\node[state,fill=gray,thick, minimum size = 3mm]         (C) [below left = 3mm of A] {};
	
	\path[->] (A) edge [bend left] node {} (B)
	(B) edge [loop right]  node (LOOP) {} (B)
		edge [bend left]   node {} (C)
	(C)	edge [bend left]  node {} (A);
	
	\node [minimum width=\mwdth, blue, minimum height=0.5cm, above = 0mm of A] (SCLABEL) {Supervisory\\Controller};
	\node[draw, thick, blue, dashed, inner sep=2mm, fit=(A) (B) (C) (LOOP) (SCLABEL)] (SC) {} ;	
	


	\draw [->, thick, color=blue] (SC) -- (Controller);
	\node [below left = \mdis/2-1mm and -0.4cm of SC] (ctrl_params) { \begin{parbox}{0.5cm}{\color{blue} Control\\Parameters}\end{parbox}};


	
	\coordinate[] (Refhelp) at ($(System) + (5*\mdis, 0)$);
	\draw[->, thick, color=blue] (Refhelp) |- (SC) node [pos = 0.7, below, color=blue] () {State Update};

	\begin{pgfonlayer}{bg1}    
	\begin{scope}[shift={(0.1,0.1)}]{
		\def\mwdth{1.25cm}
		\def\mhgth{0.5cm}
		\def\mdis{0.5cm}
		
		\node [draw, minimum width=\mwdth, minimum height=\mhgth*2, thick] (Controller1) at (0,0) {Controller} node [below = 0mm of Controller1] (Reason1) {\textit{\textbf{Reason}}};
		
		\node [draw, minimum width=\mwdth, minimum height=\mhgth, right = 3*\mdis of Controller1] (System1) {Subsystem};
		
		\node [circle, draw, minimum width=1*\mhgth, minimum height=1*\mhgth, left = \mdis of Controller1] (Diff) {};
		\node [] () at ($(Diff) - (0,0)$) {+};
		\node [] () at ($(Diff) - (0,0.25*\mhgth)$) {$-$};
		
		\draw[->,thick] (Diff) -- (Controller1);
		\draw[->,thick] (Controller1) -- (System1) node [midway, below] (Act_Freq) {\textit{\textbf{Act}}} ;
		
		\node [left= \mdis of Diff, opacity=0.0] (tmp) {Goal};
		\node (Ref1) at (tmp) {\textit{Goal}};
		\draw[->,thick] (Ref1) -- (Diff);

		\coordinate[] (FBhelp1) at ($(System1) + (5*\mdis, -2*\mdis)$);
		\draw[->,thick] (System1) -| node [near start, above] (Sen_Heart1) {\textit{\textbf{Sense}}} node [near start, below] (Sen_IPS1) {} (FBhelp) -| (Diff);
		
		\node[draw, fill=white, thick, inner sep=2mm, fit=(System1) (Controller1) (Ref1) (Sen_Heart1) (Sen_IPS1) (FBhelp1) (Reason1)] {};	
		
			\draw [->, thick, color=blue] (SC) -- (Controller1);
			\coordinate[] (Refhelp1) at ($(System1) + (5*\mdis, 0)$);
			\draw[->, thick, color=blue] (Refhelp1) |- (SC);
		
		\node[draw, fill=white, thick, inner sep=2mm, fit=(System) (Controller) (Ref) (Sen_Heart) (Sen_IPS) (FBhelp) (Reason)] {};
	}
	\end{scope}
	\end{pgfonlayer}	
	
	\begin{pgfonlayer}{bg2}    
	\begin{scope}[shift={(0.2,0.2)}]{
		\def\mwdth{1.25cm}
		\def\mhgth{0.5cm}
		\def\mdis{0.5cm}
		
		\node [draw, minimum width=\mwdth, minimum height=\mhgth*2, thick] (Controller2) at (0,0) {Controller} node [below = 0mm of Controller2] (Reason2) {\textit{\textbf{Reason}}};
		
		\node [draw, minimum width=\mwdth, minimum height=\mhgth, right = 3*\mdis of Controller2] (System2) {Subsystem};
		
		\node [circle, draw, minimum width=1*\mhgth, minimum height=1*\mhgth, left = \mdis of Controller2] (Diff) {};
		\node [] () at ($(Diff) - (0,0)$) {+};
		\node [] () at ($(Diff) - (0,0.25*\mhgth)$) {$-$};
		
		\draw[->,thick] (Diff) -- (Controller2);
		\draw[->,thick] (Controller2) -- (System2) node [midway, below] (Act_Freq) {\textit{\textbf{Act}}} ;
		
		\node [left= \mdis of Diff, opacity=0.0] (tmp) {Goal};
		\node (Ref2) at (tmp) {\textit{Goal}};
		\draw[->,thick] (Ref2) -- (Diff);
		
		\coordinate[] (FBhelp2) at ($(System2) + (5*\mdis, -2*\mdis)$);
		\draw[->,thick] (System2) -| node [near start, above] (Sen_Heart2) {\textit{\textbf{Sense}}} node [near start, below] (Sen_IPS2) {} (FBhelp2) -| (Diff);	
		
		\node[draw, fill=white, thick, inner sep=2mm, fit=(System2) (Controller2) (Ref2) (Sen_Heart2) (Sen_IPS2) (FBhelp2) (Reason2)] {};	
		
			\draw [->, thick, color=blue] (SC) -- (Controller2);
			\coordinate[] (Refhelp2) at ($(System2) + (5*\mdis, 0)$);
			\draw[->, thick, color=blue] (Refhelp2) |- (SC);
	}
	\end{scope}
    \end{pgfonlayer}

\end{tikzpicture}

%% file: fig/SPECTR_overview.tex
\begin{tikzpicture}[scale = 1.5, node distance = 1.5cm, font={\scriptsize}, >=latex, every text node part/.style={align=center}]
	\def\mwdth{2cm}
	\def\mdis{0.5cm}
	
	\node [draw, fill=white, minimum width=\mwdth, minimum height=0.5cm,] (SPLANT1) {Sub-plant 1};
	\node [draw, fill=white, minimum width=\mwdth, minimum height=0.5cm, right = \mdis of SPLANT1] (SPLANT2) {Sub-plant 2};
	\node[draw, circle, right = 1mm of SPLANT2, minimum width = 1mm, inner sep = 0mm] (DOT1)  {};
	\node[draw, circle, right = 1mm of DOT1, minimum width = 1mm, inner sep = 0mm] (DOT2)  {};
	\node[draw, circle, right = 1mm of DOT2, minimum width = 1mm, inner sep = 0mm] (DOT3)  {};
	\node [draw, fill=white, minimum width=\mwdth, minimum height=0.5cm,right = \mdis of SPLANT2] (SPLANTN) {Sub-plant \textit{N}};
	
	\begin{pgfonlayer}{bg1}
	\node[fill=gray!40, draw, dashed, thick, inner sep=2mm, fit=(SPLANT1) (SPLANT2) (SPLANTN)] (PLANT) {};
	\end{pgfonlayer}
	\node[below right] at (PLANT.south west) (SPECTRLabel) {\textbf{\color{gray}Physical Plant}};
	
	\node [draw, fill=white, minimum width=\mwdth, minimum height=0.5cm, above = 2*\mdis of SPLANT1] (CONT1) {Classic\\Controller 1};
	\node [draw, fill=white, minimum width=\mwdth, minimum height=0.5cm, right = \mdis of CONT1] (CONT2) {Classic\\Controller 2};
	\node[draw, circle, right = 1mm of CONT2, minimum width = 1mm, inner sep = 0mm] (DOT1)  {};
	\node[draw, circle, right = 1mm of DOT1, minimum width = 1mm, inner sep = 0mm] (DOT2)  {};
	\node[draw, circle, right = 1mm of DOT2, minimum width = 1mm, inner sep = 0mm] (DOT3)  {};
	\node [draw, fill=white, minimum width=\mwdth, minimum height=0.5cm,right = \mdis of CONT2] (CONTN) {Classic\\Controller \textit{N}};
	\node[draw, dashed, thick, inner sep=2mm, fit=(CONT1) (CONT2) (CONTN)] (LEAF) {};
	\node[above right] at (LEAF.north west) (SPECTRLabel) {\textbf{Leaf Controllers}};
	
	\node [draw, minimum width=\mwdth, minimum height=0.5cm, above = \mdis of LEAF] (SC) {Supervisory\\Controller};
	
	\node [draw, fill=gray!40, minimum width=\mwdth, minimum height=0.5cm, right = 2*\mdis of SC] (MODEL) {High-level\\Plant Model};
	
	\node [draw, above = \mdis of SC] (GOALS) {Variable Goals and Policies};
    \node [left= 2*\mdis of GOALS, opacity=0.0] (tmp) {};
    \draw[->,thick] (tmp) -- (GOALS) node [near start, above] (INP) {User inputs};
    
    \draw[->] (GOALS) -- (SC);
    
	\coordinate[left = 2mm of CONT1.south] (help0);
	\coordinate[left = 2mm of SPLANT1.north] (help1);
	\draw[->] (help0) -- (help1);
	\coordinate[left = 2mm of CONT2.south] (help0);
	\coordinate[left = 2mm of SPLANT2.north] (help1);
	\draw[->] (help0) -- (help1);
	\coordinate[left = 2mm of CONTN.south] (help0);
	\coordinate[left = 2mm of SPLANTN.north] (help1);
	\draw[->] (help0) -- (help1);

	\coordinate[right = 2mm of CONT1.south] (help0);
	\coordinate[right = 2mm of SPLANT1.north] (help1);
	\draw[->] (help1) -- (help0) node [midway, left] {\tiny \begin{parbox}{1.1cm}{\texttt{Commands}}\end{parbox}};
	\coordinate[right = 2mm of CONT2.south] (help0);
	\coordinate[right = 2mm of SPLANT2.north] (help1);
	\draw[->] (help1) -- (help0) node [midway, left] {\tiny \begin{parbox}{1.1cm}{\texttt{Commands}}\end{parbox}};
	\coordinate[right = 2mm of CONTN.south] (help0);
	\coordinate[right = 2mm of SPLANTN.north] (help1);
	\draw[->] (help1) -- (help0) node [midway, left] {\tiny \begin{parbox}{1.1cm}{\texttt{Commands}}\end{parbox}};

	\draw[->] (SC) -- (CONT1);
	\draw[->] (SC) -- (CONT2) node [near start, right] {\tiny \begin{parbox}{0.5cm}{ \texttt{Control\\Parameters}}\end{parbox}};
	\draw[->] (SC) -- (CONTN);

	\coordinate[above = 2mm of SC.east] (help0);
	\coordinate[above = 2mm of MODEL.west] (help1);
	\draw[->, dashed] (help0) -- (help1) node [midway, above] {\tiny \begin{parbox}{0.5cm}{ \texttt{Virtual\\Control}}\end{parbox}};

	\coordinate[below = 2mm of SC.east] (help0);
	\coordinate[below = 2mm of MODEL.west] (help1);
	\draw[->] (help1) -- (help0) node [midway, below] {\tiny \begin{parbox}{0.5cm}{ \texttt{Information}}\end{parbox}};
	
	\coordinate[right = \mdis of PLANT.east] (help0);
	\draw[->] (PLANT) -- (help0) |- (MODEL);
	
	\coordinate[below = 2mm of PLANT.east] (help1);
	\coordinate[right = 2*\mdis of help1] (help0);
	\draw[->] (help1) -- (help0) node [below] {\tiny \texttt{State Updates}} |- (GOALS) ;
	
	\node[draw, dashed, very thick, red, fit=(LEAF) (SC) (MODEL)] (SPECTR) {};
	\node[below right, color=red] at (SPECTR.north west) (SPECTRLabel) {SPECTR};

\end{tikzpicture}

%% file: fig/spectr_platform.tex
\begin{tikzpicture}[node distance = 1.5cm, font={\scriptsize}, >=latex, every text node part/.style={align=center}]
	\def\mwdth{5mm}
	\def\mdis{1mm}

	\node [draw, fill=white, minimum width=\mwdth, minimum height=0.5cm,] (C0) {};
	\node [draw, fill=white, minimum width=\mwdth, minimum height=0.5cm, right = 0mm of C0] (C1) {};
	\node [draw, fill=white, minimum width=\mwdth, minimum height=0.5cm, right = 0mm of C1] (C2) {};
	\node [draw, fill=white, minimum width=\mwdth, minimum height=0.5cm, right = 0mm of C2] (C3) {};
	
	\node [draw, fill=white, minimum width=0.8\mwdth, minimum height=0.8\mwdth, right = 
	\mdis of C3.north east, anchor= north west] (C4) {};
	\node [draw, fill=white, minimum width=0.8\mwdth, minimum height=0.8\mwdth, right = 0mm of C4] (C5) {};
	\node [draw, fill=white, minimum width=0.8\mwdth, minimum height=0.8\mwdth, right = 0mm of C5] (C6) {};
	\node [draw, fill=white, minimum width=0.8\mwdth, minimum height=0.8\mwdth, right = 0mm of C6] (C7) {};
	\node [opacity=100, minimum width=4.0\mwdth, below = \mdis of C3] (LITTLELABEL)  {Big.LITTLE};
	
	\begin{pgfonlayer}{bg1}    

		\node[fit=(C0) (C1) (C2) (C3)] (BIGCLUSTER) {};
		\node[fit=(C4) (C5) (C6) (C7)] (LITTLECLUSTER) {};

	    \node[draw, fill=gray!50, inner sep=\mdis, minimum width = 40.0\mwdth, fit=(LITTLECLUSTER) (BIGCLUSTER) (LITTLELABEL)] (PLATFORM) {};

    \end{pgfonlayer}
    
	\node[above = \mdis of PLATFORM.north west, anchor=south west, minimum width = 40.0\mwdth] (OS) {Linux};
	\node[draw, fill=white, right, inner sep = 1mm, anchor = east] at (OS.east) (SPECTRKERNEL) {\tiny \color{red} \textbf{SPECTR}};

	\begin{pgfonlayer}{bg1}    
		\node[draw, fill=gray!50,  fit=(OS) (SPECTRKERNEL), inner sep = 0mm] (SW) {};
	\end{pgfonlayer}
	
 	\node[draw, above = \mdis of SW, minimum height=\mwdth] (MW) {\textcolor{red}{\textbf{SPECTR}} \resizebox{20.0\mwdth}{!}{\input{fig/sct_gsc.tex}}};

	\node[above left = 2.0\mdis and 2.0\mdis of MW.north, anchor = south east] (QOSAPPS) {\includegraphics[height=\mwdth]{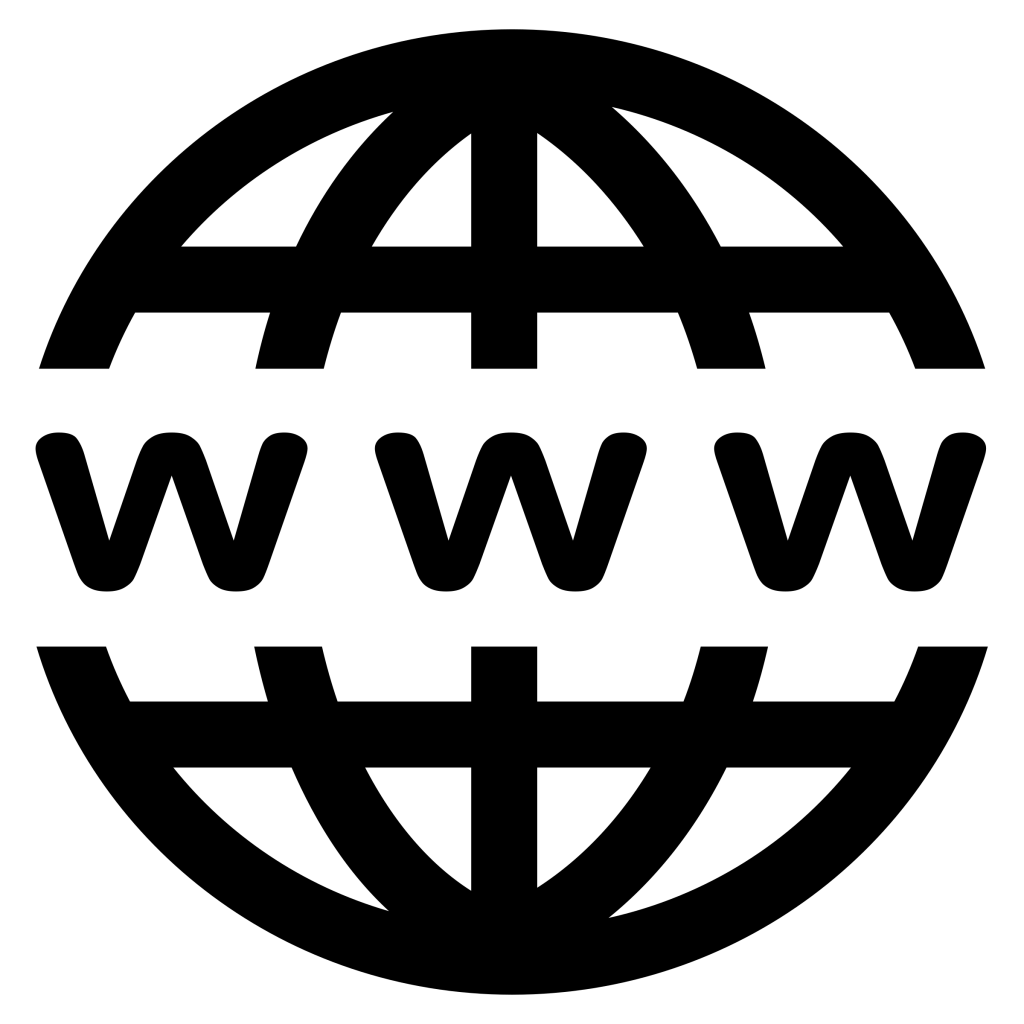}
		\includegraphics[height=\mwdth]{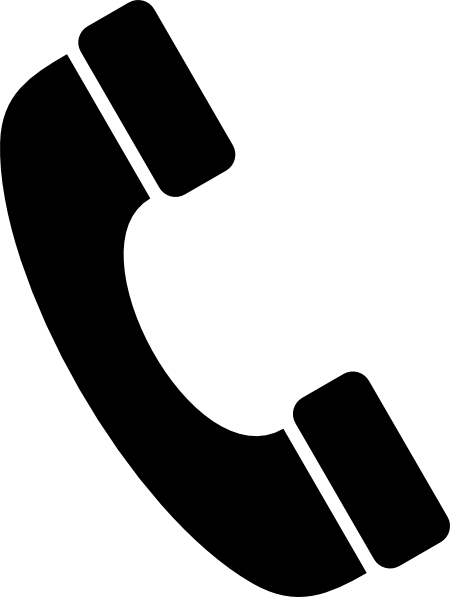}
		\includegraphics[height=\mwdth]{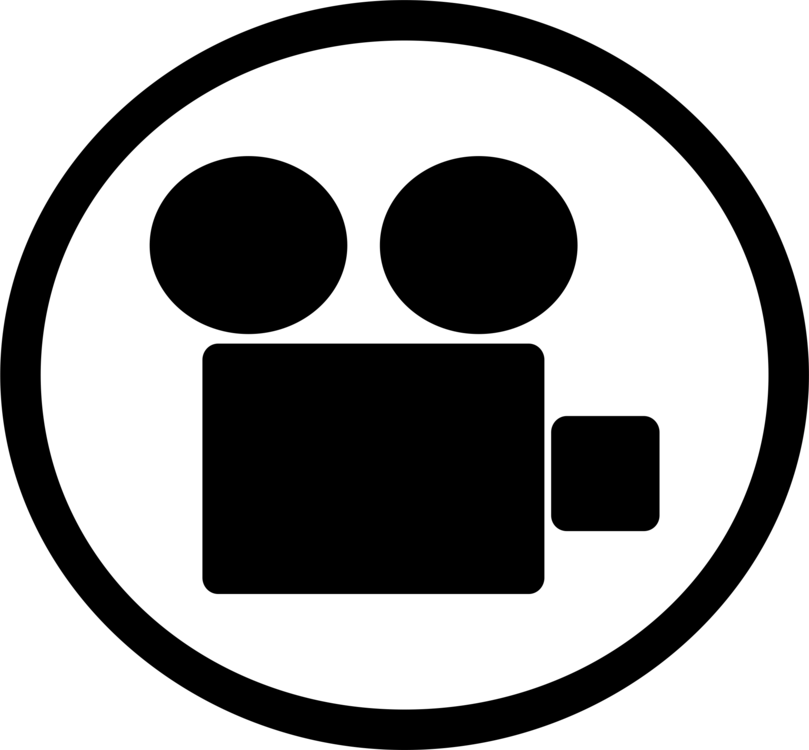}};
	\node [minimum width=3.0\mwdth, above = 0 of QOSAPPS] (QOSLABEL) {QoS};
	\node[draw, inner sep=\mdis, fit=(QOSAPPS) (QOSLABEL)] (QOSALL) {};
	
	\node[above right = 2.0\mdis and 2.0\mdis of MW.north] (NOQOSAPPS) {\includegraphics[height=\mwdth]{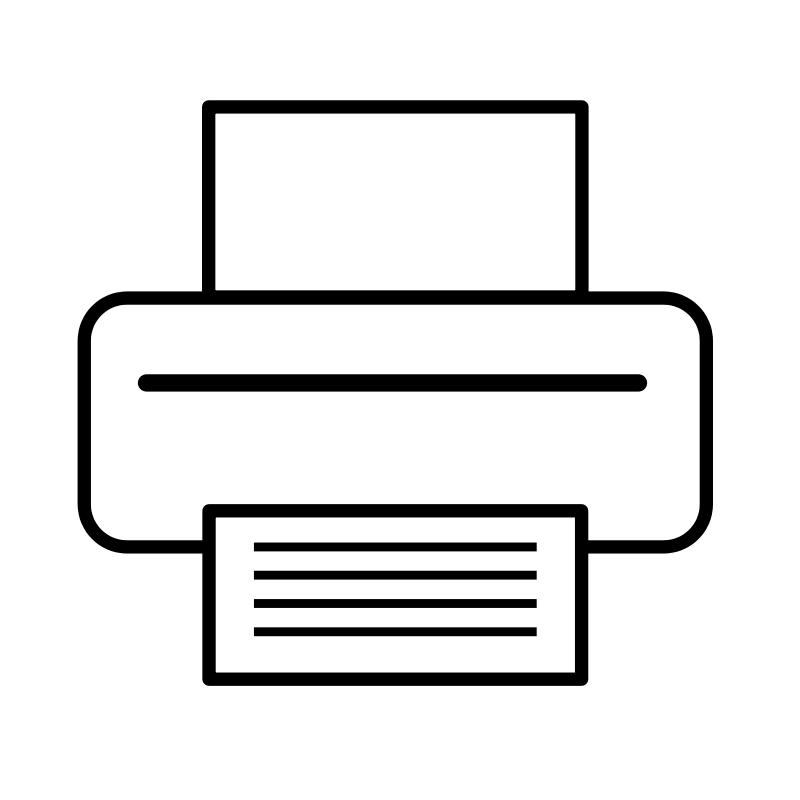}
		\includegraphics[height=\mwdth]{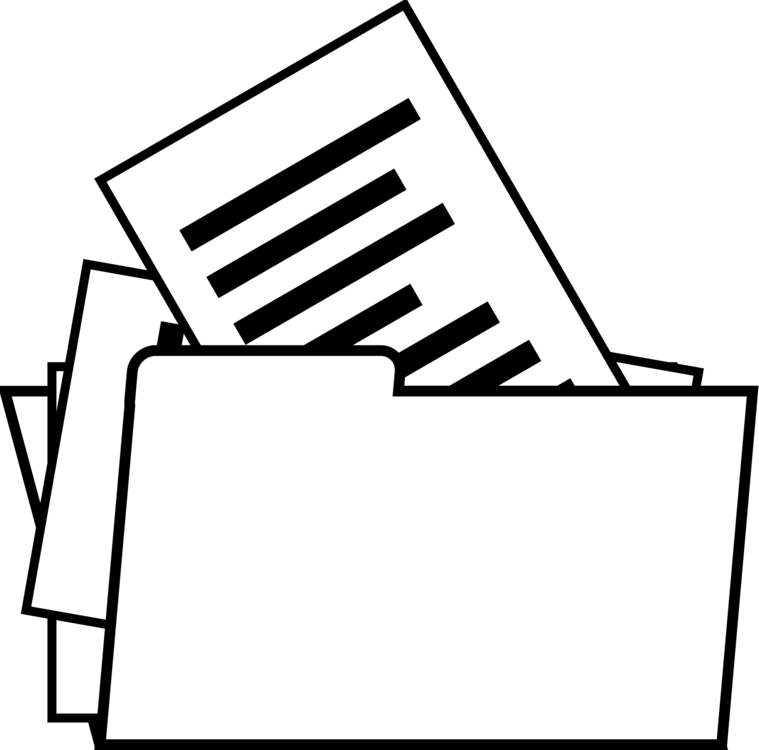}
		\includegraphics[height=\mwdth]{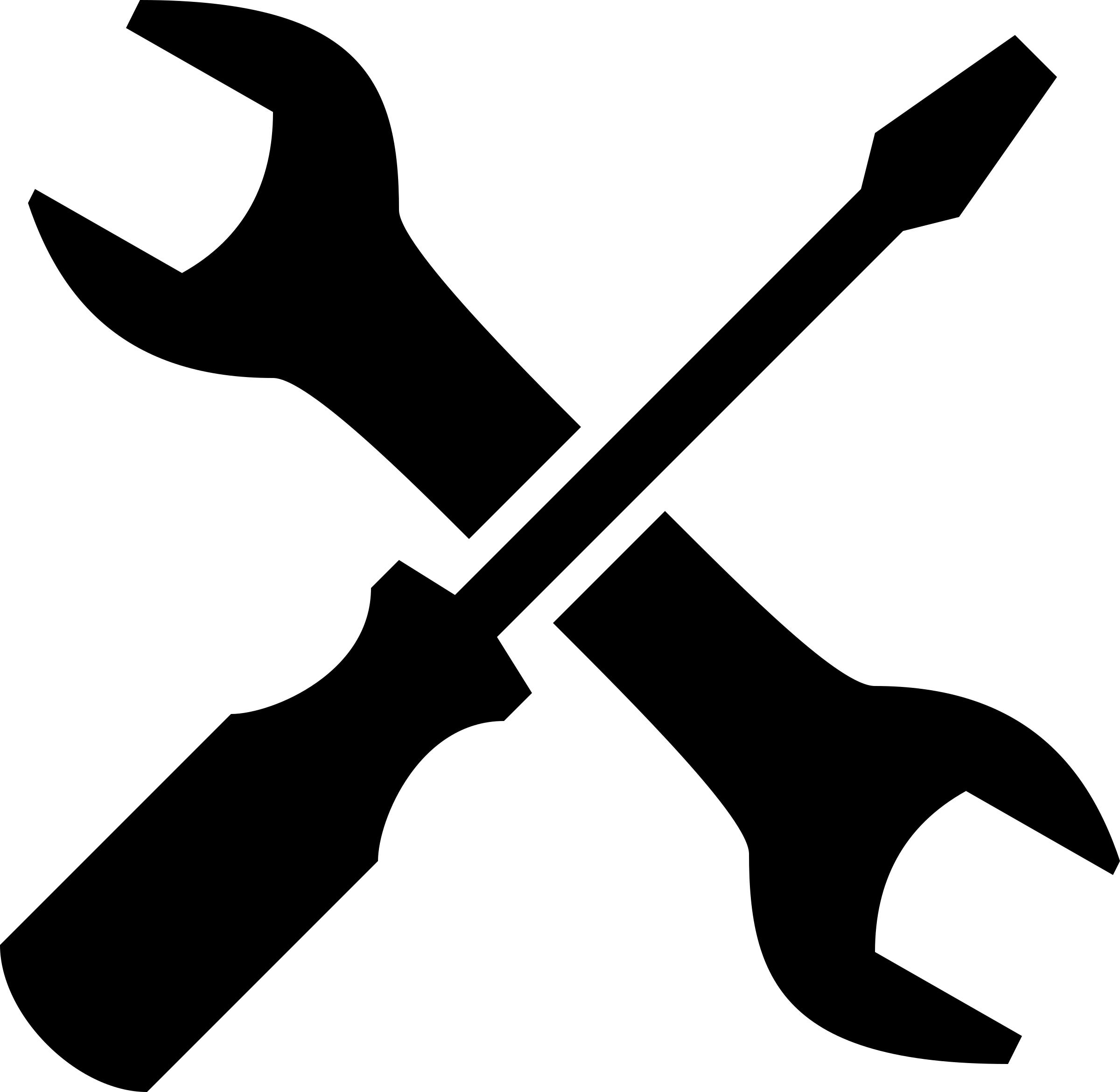}};
	\node [minimum width=3.0\mwdth, above = 0 of NOQOSAPPS] (NOQOSLABEL) {Non-QoS};
	
	\begin{pgfonlayer}{bg1}
	\node[right = 1.0\mdis of NOQOSAPPS.west] (NOQOSAPPS1) {\includegraphics[height=\mwdth]{fig/clipart-printer}
		\includegraphics[height=\mwdth]{fig/clipart-folder}
		\includegraphics[height=\mwdth]{fig/tools-clipart}};
	\node [minimum width=3.0\mwdth, above = 0 of NOQOSAPPS] (NOQOSLABEL1) {Non-QoS};
	\node[draw, inner sep=\mdis, fit=(NOQOSAPPS1) (NOQOSLABEL1), xshift=1mm, yshift=1mm] (NOQOSALL1) {};
	
	\node[draw, fill=white, inner sep=\mdis, fit=(NOQOSAPPS) (NOQOSLABEL)] (NOQOSALL) {};
	\end{pgfonlayer}
	
	\node[fit= (QOSALL) (NOQOSALL1) (NOQOSALL), inner sep = 0mm, above = \mdis of MW] () {};
	
	\node[left = \mdis of QOSALL] (APPLABEL) {Application};
	\node[below = 12.0\mdis of APPLABEL.south east, anchor = north east] (MWLABEL) {Middleware};
	\node[below = 7.0\mdis of MWLABEL.south east, anchor = north east] (OSLABEL) {Operating System};
	\node[below = 7.0\mdis of OSLABEL.south east, anchor = north east] (HWLABEL) {Hardware};
	
	\path[line width=1mm, black!30!red, ->]
	 (QOSAPPS.west) edge [bend right] node {} (MW.west)
	 (PLATFORM.west) edge [bend left] node {} (MW.west)
	 (MW.east) edge [bend left] node {} (PLATFORM.east);

\end{tikzpicture}